\documentclass[a4paper,12pt]{article}
\usepackage{amsmath,amssymb,amsthm,amsxtra,overpic,bbm,bm,epsfig,subfigure}
\usepackage{hyperref}
\usepackage{mathrsfs}
\usepackage{enumitem}
\usepackage{graphicx}
\usepackage{color}
\usepackage{comment}
\usepackage{epstopdf}
\usepackage{float}
\usepackage{cite}
\usepackage{bbold}
\usepackage{slashed,stmaryrd}
\numberwithin{equation}{section}

\def\thefootnote{\fnsymbol{footnote}}

\begin{document}

\vspace{0.2cm}

\begin{center}
{\Large\bf Density Matrix Formalism for PT-Symmetric Non-Hermitian Hamiltonians with the Lindblad Equation}
\end{center}

\vspace{0.2cm}

\begin{center}
{\bf Tommy Ohlsson}~$^{a,~b,~c}$~\footnote{E-mail: tohlsson@kth.se},
\quad
{\bf Shun Zhou}~$^{d,~e}$~\footnote{E-mail: zhoush@ihep.ac.cn}
\\
\vspace{0.2cm}
{\small $^a$Department of Physics, School of Engineering Sciences, KTH Royal Institute of Technology, \\
AlbaNova University Center, Roslagstullsbacken 21, SE-106 91 Stockholm, Sweden \\
$^b$The Oskar Klein Centre for Cosmoparticle Physics, AlbaNova University Center, \\
Roslagstullsbacken 21, SE-106 91 Stockholm, Sweden \\
$^c$University of Iceland, Science Institute, Dunhaga 3, IS-107 Reykjavik, Iceland \\
$^d$Institute of High Energy Physics, Chinese Academy of Sciences, Beijing 100049, China\\
$^e$School of Physical Sciences, University of Chinese Academy of Sciences, Beijing 100049, China}
\end{center}

\vspace{1.5cm}

\begin{abstract}
In the presence of Lindblad decoherence, i.e.~dissipative effects in an open quantum system due to interaction with an environment, we examine the transition probabilities between the eigenstates in the two-level quantum system described by non-Hermitian Hamiltonians with the Lindblad equation, for which the parity-time-reversal (PT) symmetry is conserved. First, the density matrix formalism for PT-symmetric non-Hermitian Hamiltonian systems is developed. It is shown that the Lindblad operators $L^{}_j$ are pseudo-Hermitian, namely, $\eta L^{}_j \eta^{-1} = L^\dagger_j$ with $\eta$ being a linear and positive-definite metric, and respect the PT symmetry as well. We demonstrate that the generalized density matrix $\rho^{}_{\rm G}(t) \equiv \rho(t) \eta$, instead of the normalized density matrix $\rho^{}_{\rm N}(t) \equiv \rho(t)/{\rm tr}\left[\rho(t)\right]$, should be implemented for the calculation of the transition probabilities in accordance with the linearity requirement. Second, the density matrix formalism is used to derive the transition probabilities in general cases of PT-symmetric non-Hermitian Hamiltonians. In some concrete examples, we calculate compact analytical formulas for the transition probabilities and explore their main features with numerical illustrations. We also make a comparison between our present results and our previous ones using state vectors in the absence of Lindblad decoherence.
\end{abstract}

\newpage

\def\thefootnote{\arabic{footnote}}
\setcounter{footnote}{0}

\section{Introduction}

In general, the density matrix formalism has been proven to be very useful in quantum mechanics to describe the time evolution of quantum states, no matter whether they are pure or mixed states~\cite{Weinberg:2015}. In particular, if the open quantum system under consideration interacts with an environment, the Hamiltonian of the total system can be decomposed as ${\cal H}^{}_{\rm tot} = {\cal H}^{}_{\cal S}\otimes \mathbb{1} + \mathbb{1}\otimes {\cal H}^{}_{\cal E} + g{\cal H}^\prime$, where ${\cal H}^{}_{\cal S}$ is the effective Hamiltonian of the system of our interest, ${\cal H}^{}_{\cal E}$ the Hamiltonian of the environment, and ${\cal H}^\prime$ their interaction Hamiltonian with $g$ being a small coupling constant. After tracing the density matrix of the total system $\rho_{\rm tot} = \rho_{\cal S} \otimes \rho_{\cal E}$ over the degrees of freedom of the environment, one can obtain the most general equation for the time evolution of the density matrix $\rho(t) \equiv \rho_{\cal S} = {\rm tr}_{\cal E}(\rho_{\rm tot})$ of the system in question as~\cite{Lindblad:1975ef, Gorini:1975nb}
\begin{equation}
\frac{\partial \rho}{\partial t} = -{\rm i} \left[H, \rho\right] - \frac{1}{2} \sum^{N^2 - 1}_{j = 1} \left( L^\dagger_j L^{}_j \rho + \rho L^\dagger_j L^{}_j\right) + \sum^{N^2 - 1}_{j = 1} L^{}_j \rho L^\dagger_j \,,
\label{eq:lindblad}
\end{equation}
where we have restricted ourselves to an $N$-dimensional Hilbert space $\mathfrak{h}$ with the Hamiltonian ${\cal H}^{}_{\cal S} = H$ and the density matrix $\rho(t) \equiv \sum^N_{i = 1} p^{}_i |\psi^{}_i(t)\rangle \langle \psi^{}_i(t)|$ is Hermitian by definition with $p^{}_i \geq 0$ being the probability to be in the state $|\psi^{}_i(t)\rangle$ among the complete set of Schr\"{o}dinger-picture states (for $i = 1, 2, \dots, N$) such that $\sum_{i=1}^N p^{}_i = 1$ \cite{Alicki:2007}. In the Lindblad equation in Eq.~(\ref{eq:lindblad}),\footnote{This equation was independently derived by Lindblad in Ref.~\cite{Lindblad:1975ef}, Gorini, Kossakowski, and Sudarshan in Ref.~\cite{Gorini:1975nb}, and Franke in Ref.~\cite{Franke:1976}, so it is also referred to as the Lindblad--Gorini--Kossakowski--Sudarshan--Franke equation in the literature.} there appear $N^2 - 1$ operators $L^{}_j$, which are required to be Hermitian for the von Neumann entropy to be monotonically increasing with the time evolution~\cite{Benatti:1987dz}. In addition to its Hermiticity, the trace of the density matrix $\rho(t)$ is initially normalized to one, i.e.~${\rm tr}\left[\rho(0)\right] = 1$, and it remains to be unchanged due to the unitary time evolution with Hermitian Hamiltonians $H$ in ordinary quantum mechanics~\cite{Weinberg:2015}. Obviously, the Lindblad operators $L^{}_j$ on the right-hand side of Eq.~(\ref{eq:lindblad}) should be responsible for dissipative effects in the open quantum system, i.e.~so-called Lindblad decoherence. In general, a vast variety of interesting applications of the Lindblad equation can be found in previous works, such as those applied to neutrino oscillations~\cite{Lisi:2000zt, Benatti:2000ph, Gago:2000qc, Gago:2000nv, Ohlsson:2000mj, Benatti:2001fa, Gago:2002na, Oliveira:2014jsa, Boriero:2017tkh, Coloma:2018idr, Gomes:2018inp, Gomes:2020muc}, neutron-antineutron oscillations~\cite{Kerbikov:2017spv}, and atomic clocks~\cite{Weinberg:2016uml}. In particular, in atomic, molecular, and optical physics, two-level quantum systems are fundamental and such systems investigated with the Lindblad equation consist e.g.~of two-level atoms in dissipative cavities \cite{Salah:2020}, trapped atoms \cite{Leibfried:2003zz}, qubits \cite{Parra-Lopez:2020}, and nuclear magnetic resonance \cite{Bonnard:2010} for atomic physics, molecular alignment by laser fields in dissipative media \cite{Vieillard:2008,Bonnard:2008} for molecular physics, and optical two-level systems with gain and loss (see e.g.~the review in Ref.~\cite{El-Ganainy:2018}) for optical physics.

The generalization of the density matrix formalism to non-Hermitian Hamiltonian systems has previously been extensively discussed in the literature and, for example, applied to the neutral-meson system~\cite{Benatti:1996rd, Benatti:1997rv, Benatti:1997dj, Benatti:1997xt, Bertlmann:2006fn, Bernabeu:2012au}. A general non-Hermitian Hamiltonian $H$ can always be written as $H = H^{}_+ + H^{}_-$ with $H^{}_\pm = (H \pm H^\dagger)/2$ being Hermitian ($H_+$) and anti-Hermitian ($H_-$), respectively. For the neutral-meson system~\cite{Lee:1965hi}, $H$ is usually written on the Weisskopf--Wigner form $H = M - {\rm i}\Gamma/2$, where one can immediately identify $M = H^{}_+$ and $\Gamma = 2{\rm i}H^{}_-$ and observe that both $M$ and $\Gamma$ are Hermitian. Moreover, the eigenvalues of $M$ and $\Gamma$ are restricted to be positive, since they are the physical masses and the decay widths of the mass eigenstates of the neutral mesons, respectively. In the case of non-Hermitian Hamiltonians, the Lindblad equation for the density matrix reads~\cite{Benatti:1996rd, Benatti:1997rv}
\begin{equation}
\frac{\partial \rho}{\partial t} = -{\rm i} \left(H \rho - \rho H^\dagger\right) - \frac{1}{2} \sum^{N^2 - 1}_{j = 1} \left( L^\dagger_j L^{}_j \rho + \rho L^\dagger_j L^{}_j\right) + \sum^{N^2 - 1}_{j = 1} L^{}_j \rho L^\dagger_j \,, \label{eq:Nlindblad}
\end{equation}
where the trace of the density matrix is no longer conserved. For this reason, the normalized density matrix $\rho^{}_{\rm N}(t) \equiv \rho(t)/{\rm tr}\left[\rho(t)\right]$ has been defined in Refs.~\cite{Sergi:2013, Sergi:2014eja, Ju:2019kso}, leading to the average of an observable ${\cal O}(t)$ given by $\langle {\cal O}(t) \rangle \equiv {\rm tr}\left[\rho^{}_{\rm N}(t){\cal O}(t = 0)\right]$. On the other hand, the Lindblad equation with $\eta$-pseudo-Hermitian Hamiltonians~\cite{Mostafazadeh:2008pw}, for which $\eta H \eta^{-1} = H^\dagger$ with $\eta$ being a Hermitian metric operator in the Hilbert space $\mathfrak{h}$, has been derived in Refs.~\cite{Scolarici:2006, Scolarici:2007} for the generalized density matrix $\rho^{}_{\rm G}(t) \equiv \rho(t)\eta$, viz.,
\begin{equation}
\frac{\partial \rho^{}_{\rm G}}{\partial t} = -{\rm i} \left[H, \rho^{}_{\rm G}\right] - \frac{1}{2} \sum^{N^2 - 1}_{j = 1} \left( L^\ddag_j L^{}_j \rho^{}_{\rm G} + \rho^{}_{\rm G} L^\ddag_j L^{}_j\right) + \sum^{N^2 - 1}_{j = 1} L^{}_j \rho^{}_{\rm G} L^\ddag_j \,,
\label{eq:Glindblad}
\end{equation}
where $L^\ddag_j \equiv \eta^{-1} L^\dagger_j \eta$ is defined and symbolizes the adjoint of $L_j$ with respect to the pseudo-inner product constructed by $\eta$~\cite{Scolarici:2006, Scolarici:2007}. More explicitly, Eq.~(\ref{eq:Glindblad}) has been found by first converting the $\eta$-pseudo-Hermitian Hamiltonian system via a similarity transformation into its Hermitian counterpart, to which the Lindblad equation in Eq.~(\ref{eq:lindblad}) is well applicable. Then, Eq.~(\ref{eq:Glindblad}) was obtained by transforming back to the original basis.

In the present work, we aim to explore the basic properties of the Lindblad equation in Eq.~(\ref{eq:Glindblad}) for non-Hermitian Hamiltonians with the parity-time-reversal (PT) symmetry~\cite{Bender:1998ke, Bender:1998gh, Bender:2002vv, Bender:2002yp, Bender:2003gu, Bender:2007nj, Ohlsson:2015xsa}. Taking the two-level system as an example~\cite{Bender:2002vv, Bender:2002yp}, we explicitly derive the transition probabilities between the eigenstates in the density matrix formalism. The main motivation for such an investigation is two-fold. First, it must be helpful to examine non-Hermitian Hamiltonians with non-trivial symmetries, which turn out to be the PT symmetry in our case. As we will see later, the Lindblad operators will be further constrained to be pseudo-Hermitian, implying a smaller number of free parameters for dissipative effects. Second, the transition probabilities between the eigenstates in the PT-symmetric non-Hermitian Hamiltonian system have been computed for the first time in Refs.~\cite{Ohlsson:2019noy, Ohlsson:2020idi}, so it is interesting to perform the calculations in the density matrix formalism and make a comparison between the results by the two different approaches. Such a comparison in some concrete examples can be achieved by just switching off the Lindblad operators. Applications of two-level quantum systems with the Lindblad equation for PT-symmetric non-Hermitian Hamiltonians have been discussed in the literature. For example, a PT-symmetric optical lattice would be a prototypical system that exhibits the characteristics of a two-level quantum system, see the review in Ref.~\cite{El-Ganainy:2018}, and other examples of two-level dissipative quantum systems have been investigated in Refs.~\cite{Cherbal:2006,Sugny:2007}.

The remaining part of this work is organized as follows. In Sec.~\ref{sec:formalism}, we first give the general formalism on the two-level quantum system described by PT-symmetric non-Hermitian Hamiltonians, and then derive the Lindblad equations for the generalized and normalized density matrices in detail. It is pointed out that the time evolution of the normalized density matrix $\rho^{}_{\rm N}(t)$ involves a non-linear term, which runs into contradiction with the linearity requirement for deriving the Lindblad equation. In Sec.~\ref{sec:prob}, the derivation of the transition probabilities between the eigenstates using the generalized density matrix is performed, where both analytical and numerical results are presented. In Sec.~\ref{sec:summary}, we summarize our main results and conclude. Finally, in Appendices~\ref{app:lindbladterm} and \ref{app:4Drep}, some useful details on the derivations of the Lindblad equations are collected, and in Appendix~\ref{app:procedure}, the procedure for our calculation of the transition probabilities is outlined for reference.

\section{General Formalism}
\label{sec:formalism}

\subsection{Two-Level System}

Non-Hermitian Hamiltonians with PT symmetry have attracted a lot of attention in recent years~\cite{Bender:2007nj} and plenty of applications have been found in many research areas other than particle physics~\cite{El-Ganainy:2018}. In the introduction, we have presented several applications of two-level quantum systems in atomic, molecular, and optical physics that can be described by non-Hermitian Hamiltonians with PT-symmetry. For example, non-Hermitian Hamiltonians have been investigated in Refs.~\cite{Cherbal:2006,Salah:2020} and several such Hamiltonians with PT symmetry for different physical systems (based on both theoretical and experimental studies) have been reviewed in Ref.~\cite{El-Ganainy:2018}. For brevity, we recall some necessary concepts of a PT-symmetric non-Hermitian Hamiltonian for a two-level quantum system~\cite{Bender:2002yp,Bender:2002vv}. One can find a detailed account of the properties of the two-level system with PT-symmetric non-Hermitian Hamiltonians in Refs.~\cite{Ohlsson:2019noy,Ohlsson:2020idi}. The most general form of such a Hamiltonian $H$ is conventionally parametrized in terms of four real parameters $r$, $s$, $\varphi$, and $\phi$, i.e.
\begin{equation}
H = \left( \begin{matrix} r e^{{\rm i} \varphi} & s e^{{\rm i} \phi} \cr s e^{-{\rm i} \phi} & r e^{-{\rm i} \varphi} \end{matrix} \right)
\label{eq:H}
\end{equation}
with two eigenvalues
$$
\lambda^{}_\pm = r \cos \varphi \pm \sqrt{s^2 - r^2 \sin^2 \varphi} \,,
$$
which are independent of the phase $\phi$. In the PT-symmetric phase with $s^2 > r^2\sin^2\varphi$, which is always assumed in the following discussion, we have two real eigenvalues $\lambda^{}_+$ and $\lambda^{}_-$. Assuming that $\phi = 0$ and following the conventions in Ref.~\cite{Ohlsson:2019noy}, one can construct two properly normalized eigenvectors as
\begin{equation}
|u^{}_\pm\rangle = \frac{1}{\sqrt{2\cos\alpha}} \left(\begin{matrix} e^{\pm{\rm i}\alpha/2} \cr \pm e^{\mp {\rm i}\alpha/2}\end{matrix}\right) \,,
\label{eq:neigenv}
\end{equation}
where $\sin \alpha \equiv r \sin\varphi/s$. Note that it is not a strong constraint to assume that $\phi = 0$, since the two eigenvalues $\lambda^{}_\pm$, i.e.~the eigenvalues of $H$ in Eq.~(\ref{eq:H}), are independent of $\phi$. Furthermore, using these two normalized eigenvectors, one can introduce the metric operator $\eta$, as shown in Ref.~\cite{Kleefeld:2009vd}, viz.,
\begin{equation}
\eta \equiv \left( |u^{}_+\rangle \langle u^{}_+| + |u^{}_-\rangle \langle u^{}_-| \right)^{-1} = \left( \begin{matrix} \sec \alpha & -{\rm i} \tan \alpha \cr {\rm i} \tan \alpha & \sec \alpha \end{matrix} \right) \,.
\label{eq:metric}
\end{equation}
The metric $\eta$ is positive-definite with determinant $\det \eta = \sec^2\alpha - \tan^2\alpha = 1$. Note that the PT symmetry of the non-Hermitian Hamiltonian $H$ manifests itself via the metric $\eta$, since it depends on the particular choices of the ${\cal P}$ and ${\cal T}$ operators (see e.g.~Ref.~\cite{Ohlsson:2019noy}). In our investigation, we assume a representation matrix of the parity operator ${\cal P}$ as
\begin{eqnarray}
{\cal P} = \left( \begin{matrix} 0 & 1 \cr 1 & 0 \end{matrix}\right) \,,
\end{eqnarray}
while the time-reversal operator ${\cal T}$ can be regarded as the combination of any unitary operator and the ordinary complex conjugate ${\cal K}$. For a PT-symmetric non-Hermitian Hamiltonian $H$, it holds that $\eta H \eta^{-1} = H^\dagger$ and $[H, {\cal PT}] = \mathbb{0}^{}_2$. In the special case that $s = 0$, i.e.~the Hamiltonian is diagonal with eigenvalues $\lambda^{}_\pm = r e^{\pm {\rm i} \varphi}$, we find that the normalized eigenvectors and the metric are
\begin{equation}
|u^r_+\rangle = \left( \begin{matrix} 1 \cr 0 \end{matrix} \right) \,, \quad |u^r_-\rangle = \left( \begin{matrix} 0 \cr 1 \end{matrix} \right) \,, \quad \eta^{}_r = \left( \begin{matrix} 1 & 0 \cr 0 & 1 \end{matrix} \right) = \mathbb{1}^{}_2 \,,
\label{eq:eigenvmetricr}
\end{equation}
whereas in the special case that $r = 0$ (or $\varphi = 0$), with the help of Eq.~(\ref{eq:neigenv}), we simply have
\begin{equation}
|u^s_+\rangle = \frac{1}{\sqrt{2}} \left( \begin{matrix} 1 \cr 1 \end{matrix} \right) \,, \quad |u^s_-\rangle = \frac{1}{\sqrt{2}} \left( \begin{matrix} 1 \cr -1 \end{matrix} \right) \,, \quad \eta^{}_s = \left( \begin{matrix} 1 & 0 \cr 0 & 1 \end{matrix} \right) = \mathbb{1}^{}_2 \,.
\end{equation}
It is worthwhile to point out the essential difference between these two cases. The Hamiltonian with $s = 0$ is still non-Hermitian, although diagonal, whereas the one with $r = 0$ (or $\varphi = 0$) is actually Hermitian. These two simple cases will be further examined later on in our investigation, as illustrative examples for the calculation of transition probabilities.

\subsection{Generalized Density Matrix}

As shown in Ref.~\cite{Ohlsson:2019noy}, for the PT-symmetric non-Hermitian Hamiltonian in Eq.~(\ref{eq:H}) with $\phi = 0$, one can find the following Hermitian matrix
\begin{equation}
G = \frac{1}{\sqrt{\cos\alpha}} \left( \begin{matrix}
  \cos \tfrac{\alpha}{2} & -{\rm i} \sin \tfrac{\alpha}{2} \cr
  {\rm i} \sin \tfrac{\alpha}{2} & \cos \tfrac{\alpha}{2}
\end{matrix} \right)
\label{eq:G}
\end{equation}
such that the transformed Hamiltonian $H^\prime \equiv G H G^{-1}$ is Hermitian, i.e.
\begin{equation}
H^\prime = \left( \begin{matrix} r \cos\varphi & \sqrt{s^2 - r^2 \sin^2 \varphi} \cr \sqrt{s^2 - r^2 \sin^2 \varphi} & r \cos\varphi \end{matrix} \right) \,.
\end{equation}
Note that $G^2 = \eta$ and $G^{-1} = G^* = G^{\rm T}$ hold. In such a Hermitian basis, the state vector $|\psi^\prime(t)\rangle$ is related to the original one via $|\psi^\prime(t)\rangle = G |\psi^{}(t)\rangle$, indicating that $\rho^\prime(t) = G \rho(t) G = G \rho^{}_{\rm G}(t) G^{-1}$, where the generalized density matrix $\rho^{}_{\rm G}(t) \equiv \rho(t) \eta$ has been defined. Hence, the Lindblad equation in the Hermitian basis takes on the form in Eq.~(\ref{eq:lindblad}) with $N = 2$ as in ordinary Hermitian quantum mechanics, and can be rewritten as
\begin{equation}
\frac{\partial \rho^\prime(t)}{\partial t} = -{\rm i} \left[ H^\prime,   \rho^\prime(t) \right] - \frac{1}{2} \sum^3_{j=1} \left[ L^{\prime \dagger}_j L^\prime_j \rho^\prime(t) + \rho^\prime(t) L^{\prime \dagger}_j L^\prime_j \right] + \sum^3_{j=1} L^\prime_j \rho^\prime(t) L^{\prime \dagger}_j \,,
\label{eq:hermitianlind}
\end{equation}
where the density matrix $\rho^\prime(t)$ is Hermitian and $L^\prime_j$ denote the Lindblad operators that should be Hermitian under the requirement of complete positivity and increasing von Neumann entropy~\cite{Benatti:1987dz}. To derive the Lindblad equation for the generalized density matrix $\rho^{}_{\rm G}(t)$, given $G^{-1} \rho^\prime(t) G = \rho^{}_{\rm G}(t)$, we multiply Eq.~(\ref{eq:hermitianlind}) by $G^{-1}$ from the left and $G$ from the right, and then arrive at
\begin{equation}
\frac{\partial \rho^{}_{\rm G}(t)}{\partial t} = -{\rm i} \left[ H, \rho^{}_{\rm G}(t) \right] - \frac{1}{2} \sum^3_{j=1} \left[ L^2_j \rho^{}_{\rm G}(t) + \rho^{}_{\rm G}(t) L^2_j \right] + \sum^3_{j=1} L^{}_j \rho^{}_{\rm G}(t) L^{}_j \,,
\label{eq:Glindbladh}
\end{equation}
where $L^{}_j \equiv G^{-1} L^{\prime}_j G$ have been defined. In contrast to the Hermitian Lindblad operator $L^{\prime \dagger}_j = L^\prime_j$ in the Hermitian basis, we now have $L^\dagger_j = G L^\prime_j G^{-1} = G^2 L^{}_j G^{-2} = \eta L^{}_j \eta^{-1}$ that is $\eta$-pseudo-Hermitian, namely, $L^\ddag_j = L^{}_j$. This feature has not been noticed in Refs.~\cite{Scolarici:2006, Scolarici:2007}. It is worthwhile to stress that the Lindblad equation in Eq.~(\ref{eq:Glindbladh}) for $\rho^{}_{\rm G}(t)$ and $L^{}_j$ resembles exactly the same form in Eq.~(\ref{eq:lindblad}) for $\rho(t)$ and $L^{}_j$ in ordinary Hermitian quantum mechanics. However, both $\rho(t)$ and $L^{}_j$ are Hermitian in the latter case, whereas $\rho^{}_{\rm G}(t)$ and $L^{}_j$ are $\eta$-pseudo-Hermitian in the former. It should be emphasized that $L_j$ are responsible for dissipative effects in the two-level quantum systems under investigation and have to be identified for each given such system. It is interesting to observe if the general Lindblad equation for PT-symmetric non-Hermitian Hamiltonians can be derived by extending the simple strategy outlined in Ref.~\cite{Pearle:2012} in ordinary quantum mechanics with Hermitian Hamiltonians. In applications of two-level quantum systems with dissipative effects, there are many examples of typical systems of interest that can be investigated. For example, in atomic, molecular, and optical physics, interesting systems consist of two-level atoms in dissipative cavities~\cite{Salah:2020}, dissipative two-level spin systems in nuclear magnetic resonance~\cite{Cherbal:2006}, molecular alignment in dissipative media~\cite{Vieillard:2008,Bonnard:2008}, and a PT-symmetric optical lattice as a two-level quantum system with gain and loss \cite{El-Ganainy:2018}.

\subsection{Normalized Density Matrix}

Similarly to our derivation of the Lindblad equation for $\rho^{}_{\rm G}(t)$, we now study the evolution equation for the normalized density matrix $\rho^{}_{\rm N}(t) \equiv \rho(t)/{\rm tr}\left[\rho(t)\right]$. First of all, one has to establish the relationship between $\rho^{}_{\rm N}(t)$ and the density matrix $\rho^\prime(t)$ in the Hermitian basis. Starting with the relation $\rho^\prime(t) = G \rho(t) G$, we obtain
\begin{equation}
{\rm tr}\left[\rho(t)\right] = {\rm tr}\left[G^{-1} \rho^\prime(t) G^{-1}\right] = {\rm tr} \left[\eta^{-1} \rho^\prime(t)\right] \,,
\label{eq:trace}
\end{equation}
leading to
\begin{align}
\rho^{}_{\rm N}(t) &= \frac{G^{-1} \rho^\prime(t) G^{-1}}{{\rm tr} \left[\eta^{-1} \rho^\prime(t)\right]} \,, \label{eq:rhoNh1} \\
\frac{\partial \rho^{}_{\rm N}(t)}{\partial t} &= \frac{G^{-1} \left[\partial \rho^\prime(t)/\partial t \right] G^{-1}}{{\rm tr} \left[\eta^{-1} \rho^\prime(t)\right]} - \rho^{}_{\rm N}(t) \frac{{\rm tr} \left[\eta^{-1} \partial \rho^\prime(t)/\partial t \right]}{{\rm tr} \left[\eta^{-1} \rho^\prime(t)\right]} \,. \label{eq:rhoNh2}
\end{align}
Then, from the Lindblad equation for $\rho^\prime(t)$ in Eq.~(\ref{eq:hermitianlind}), one can derive the Lindblad equation for $\rho^{}_{\rm N}(t)$ as follows
\begin{equation}
\frac{\partial \rho^{}_{\rm N}(t)}{\partial t} = -{\rm i} \left[H \rho^{}_{\rm N}(t) - \rho^{}_{\rm N}(t) H^\dagger\right] - \frac{1}{2} \sum^3_{j=1} \left[ L^2_j \rho^{}_{\rm N}(t) + \rho^{}_{\rm N}(t) L^{\dagger 2}_j \right] + \sum^3_{j=1} L^{}_j \rho^{}_{\rm N}(t) L^{\dagger}_j + {\cal N}\left[\rho^{}_{\rm N}(t)\right] \,,
\label{eq:Nlindbladh}
\end{equation}
where $L^{}_j \equiv G^{-1} L^{\prime}_j G$ have been defined in the exactly same way as in Eq.~(\ref{eq:Glindbladh}) and the last term on the right-hand side is given by
\begin{equation}
{\cal N}\left[\rho^{}_{\rm N}(t)\right] = 2{\rm i}\, {\rm tr}\left[H^{}_- \rho^{}_{\rm N}(t)\right] \rho^{}_{\rm N}(t) + \frac{1}{2} \sum^3_{j=1} {\rm tr} \left[ L^2_j \rho^{}_{\rm N}(t) + \rho^{}_{\rm N}(t) L^{\dagger 2}_j \right] \rho^{}_{\rm N}(t) - \sum^3_{j=1} {\rm tr} \left[L^{}_j \rho^{}_{\rm N}(t) L^{\dagger}_j \right] \rho^{}_{\rm N}(t) \,.
\label{eq:N}
\end{equation}
It should be observed that the first term on the right-hand side of Eq.~(\ref{eq:N}) has been previously derived in Ref.~\cite{Sergi:2013}, but in a scenario without the Lindblad term. For our investigation, one should note that
\begin{equation}
- \frac{1}{2} \sum^3_{j=1} {\rm tr} \left[ L^2_j \rho^{}_{\rm N}(t) + \rho^{}_{\rm N}(t) L^{2}_j \right] + \sum^3_{j=1} {\rm tr} \left[L^{}_j \rho^{}_{\rm N}(t) L^{}_j \right] = 0 \,,
\end{equation}
implying that the contribution from the Lindblad term to the second term on the right-hand side of Eq.~(\ref{eq:rhoNh2}) vanishes only if $L^\dagger_j = L^{}_j$ holds.

\subsection{Evolution Equations}
\label{sub:evoleq}

As demonstrated in Ref.~\cite{Benatti:1996rd}, it is convenient to expand the density matrix $\rho^{}_{\rm G}(t)$ or $\rho^{}_{\rm N}(t)$, the non-Hermitian Hamiltonian $H$, and the Lindblad operators $L^{}_j$ (for $j = 1, 2, 3$) in terms of the Pauli matrices ${\bm \sigma} \equiv (\sigma_1,\sigma_2,\sigma_3) = (\sigma_i)$ and the identity matrix $\sigma^{}_0 \equiv \mathbb{1}_2$. In the following, the two cases of the generalized and normalized density matrices will be discussed in parallel.
\begin{itemize}
\item {\it The generalized density matrix} --- If we expand the relevant matrices as stated above, then we have
\begin{align}
\rho^{}_{\rm G} &= \rho^{}_\mu \sigma^{}_\mu = \rho^{}_0 \sigma^{}_0 + {\bm \rho} \cdot {\bm \sigma} \,, \\
H &= H^{}_\mu \sigma^{}_\mu = H^{}_0 \sigma^{}_0 + {\bm H}\cdot {\bm \sigma} \,, \\
L^{}_j &= L^\mu_j \sigma^{}_\mu = L^0_j \sigma^{}_0 + {\bm L}^{}_j \cdot {\bm \sigma} \,,
\end{align}
where all dependence on time $t$ has been suppressed and the corresponding coefficients $\rho^{}_\mu$, $H^{}_\mu$ and $L^\mu_j$ are in general complex. The Lindblad term defined as
\begin{equation}
{\cal L}\left[\rho^{}_{\rm G}(t)\right] \equiv - \frac{1}{2} \sum^3_{j=1} \left[ L^2_j \rho^{}_{\rm G}(t) + \rho^{}_{\rm G}(t) L^2_j \right] + \sum^3_{j=1} L^{}_j \rho^{}_{\rm G}(t) L^{}_j \,,
\end{equation}
is found to be
\begin{equation}
{\cal L}[\rho^{}_{\rm G}(t)] =-2\sum^3_{j=1} \bm{L}^2_j (\bm{\rho}\cdot \bm{\sigma}) + 2\sum^3_{j=1} (\bm{\rho}\cdot \bm{L}^{}_j) (\bm{L}^{}_j \cdot \bm{\sigma}) \,,
\end{equation}
for which the details of derivation have been collected in Appendix~\ref{app:lindbladterm}. As a consequence, the Lindblad equation for the generalized density matrix $\rho_{\rm G}(t)$ in Eq.~(\ref{eq:Glindbladh}) turns out to be
\begin{equation}
\frac{\partial \rho^{}_{\rm G}(t)}{\partial t} = \frac{\partial}{\partial t} \left(\rho^{}_0 \sigma^{}_0 + {\bm \rho} \cdot {\bm \sigma}\right) = 2 \left({\bm H}\times {\bm \rho}\right) \cdot {\bm \sigma} - 2\sum^3_{j=1} \bm{L}^2_j (\bm{\rho}\cdot \bm{\sigma}) + 2\sum^3_{j=1} (\bm{\rho}\cdot \bm{L}^{}_j) (\bm{L}^{}_j \cdot \bm{\sigma}) \,.
\label{eq:Gsigma}
\end{equation}
Some helpful comments on the above equation are in order.
\begin{itemize}
\item Taking the trace on both sides of Eq.~(\ref{eq:Gsigma}), we can observe that the coefficient $\rho^{}_0(t)$ is actually constant, i.e.~$\partial \rho^{}_0(t)/\partial t = 0$. Therefore, if the generalized density matrix $\rho^{}_{\rm G}(t)$ is initially normalized as ${\rm tr} \left[\rho^{}_{\rm G}(0)\right] = 2\rho^{}_0(0) = 1$, then we have ${\rm tr}\left[\rho^{}_{\rm G}(t)\right] = 1$ for all times $t$.

\item Taking the trace on both sides of Eq.~(\ref{eq:Gsigma}) together with the Pauli matrix $\sigma^{}_i$, we obtain
\begin{equation}
\frac{\partial}{\partial t} \rho^{}_i = 2 \epsilon^{}_{ijk} H^{}_j \rho^{}_k - 2 \sum^3_{j=1} {\bm L}^2_j \rho^{}_i  + 2 \sum^3_{j=1}  (\bm{\rho}\cdot \bm{L}^{}_j) L^i_j \,,
\end{equation}
which can be recast into matrix form as
\begin{equation}
\frac{\partial}{\partial t} \left( \begin{matrix} \rho^{}_1 \cr \rho^{}_2 \cr \rho^{}_3 \end{matrix} \right) = -2 \left( \begin{matrix} A & D + H^{}_3 & E - H^{}_2 \cr D - H^{}_3 & B & F + H^{}_1 \cr E + H^{}_2 & F - H^{}_1 & C \end{matrix} \right) \left( \begin{matrix} \rho^{}_1 \cr \rho^{}_2 \cr \rho^{}_3 \end{matrix} \right)\label{eq:Glindblad3}
\end{equation}
with $H^{}_1 \equiv s \cos\phi$, $H^{}_2 \equiv - s \sin\phi$, $H^{}_3 \equiv {\rm i}r \sin\varphi$ as well as
\begin{align}
A &\equiv \sum^3_{j=1} \bm{L}^2_j - \sum_{j=1}^3 L^1_j L^1_j = \sum_{j=1}^3 \left[(L^2_j)^2 + (L^3_j)^2\right] \,, \label{eq:A}\\
B &\equiv \sum^3_{j=1} \bm{L}^2_j - \sum_{j=1}^3 L^2_j L^2_j = \sum_{j=1}^3 \left[(L^1_j)^2 + (L^3_j)^2\right] \,, \\
C &\equiv \sum^3_{j=1} \bm{L}^2_j - \sum_{j=1}^3 L^3_j L^3_j = \sum_{j=1}^3 \left[(L^1_j)^2 + (L^2_j)^2\right] \,, \\
D &\equiv - \sum^3_{j=1} L^1_j L^2_j \,, \\
E &\equiv - \sum^3_{j=1} L^1_j L^3_j \,, \\
F &\equiv - \sum^3_{j=1} L^2_j L^3_j \,. \label{eq:F}
\end{align}
Note that the Lindblad operators $L^{}_j$ are no longer Hermitian, so the coefficients $L^i_j$ are not necessarily real. Compared to the four-dimensional representation in Appendix~\ref{app:4Drep} in Eq.~(\ref{eq:Glindblad4}), we have reduced the Lindblad equation to a three-dimensional representation in Eq.~(\ref{eq:Glindblad3}) by removing the zeroth row and column related to $\rho^{}_0$.
\end{itemize}

\item {\it The normalized density matrix} --- In a similar way, we also expand the normalized density matrix $\rho^{}_{\rm N}(t)$ in terms of $\sigma^{}_\mu$ (for $\mu = 0, 1, 2, 3$) as
\begin{equation}
\rho^{}_{\rm N} = \rho^{}_\mu \sigma^{}_\mu = \rho^{}_0 \sigma^{}_0 + {\bm \rho} \cdot {\bm \sigma} \,,
\end{equation}
whereas the expansions of the non-Hermitian Hamiltonian and the Lindblad operators are identical with those in the previous case. Although the expansion is now performed for the normalized density matrix $\rho^{}_{\rm N}(t)$, we have used the same notation of the relevant coefficients $\rho^{}_\mu$, which will be clarified whenever there may occur a confusion.

The expansion of the Lindblad term remains the same, so we focus on the other terms. As one can observe in Appendix~\ref{app:4Drep}, the four-dimensional representation of the Lindblad equation for the normalized density matrix $\rho^{}_{\rm N}(t)$ has been derived for the most general non-Hermitian Hamiltonian. Now, we apply it to the PT-symmetric Hamiltonian in Eq.~(\ref{eq:H}). First, we should note that the expansion of the Hamiltonian in Eq.~(\ref{eq:H}) gives rise to $H^{}_0 = r\cos\varphi$, $H^{}_1 = s\cos\phi$, $H^{}_2 = - s\sin\phi$ and $H^{}_3 = {\rm i}r\sin\varphi$, where only $H^{}_3$ is complex and has a non-vanishing imaginary part. In this case, we obtain
\begin{equation}
\frac{\partial}{\partial t} \left( \begin{matrix} \rho^{}_0 \cr \rho^{}_1 \cr \rho^{}_2 \cr \rho^{}_3 \end{matrix} \right) = -2 \left(\begin{matrix} 0 & 0 & 0 & -r\sin\varphi \cr 0 & A & D & E + s\sin\phi \cr 0 & D & B & F + s\cos\phi \cr -r\sin\varphi & E-s\sin\phi & F-s\cos\phi & C \end{matrix}\right) \left( \begin{matrix} \rho^{}_0 \cr \rho^{}_1 \cr \rho^{}_2 \cr \rho^{}_3 \end{matrix} \right)  \,, \quad
\label{eq:Nlindblad4m}
\end{equation}
where the non-linear term has been removed compared to Eq.~(\ref{eq:Nlindblad4}). It is evident that $\rho^{}_0(t)$ itself is no longer constant in time, which is inconsistent with the basic property of the normalized density matrix, i.e.~${\rm tr}\left[\rho^{}_{\rm N}(t)\right] = 2\rho^{}_0(t) = 1$. The reason is simply the omission of the non-linear term, in which the time evolution of $\rho^{}_0(t)$ is entangled with that of $\rho^{}_i(t)$ (for $i = 1, 2, 3$). Therefore, the non-linear term in Eq.~(\ref{eq:N}) is necessary to guarantee the normalization condition, but it will be problematic to require that the Lindblad term should be linear in the normalized density matrix.
\end{itemize}

As mentioned, the Lindblad operators $L^{}_j$ (for $j = 1, 2, 3$) have to satisfy the condition of $\eta$-pseudo-Hermiticity, i.e.~$L^\dagger_j = \eta L^{}_j \eta^{-1}$. To explore the constraint on the Lindblad operators from pseudo-Hermiticity, we first assume the most general form
\begin{equation}
L^{}_j = \left( \begin{matrix} a^{}_j & b^{}_j \cr c^{}_j & d^{}_j \end{matrix} \right) \,,
\end{equation}
where $a^{}_j$, $b^{}_j$, $c^{}_j$, and $d^{}_j$ are all complex parameters. Then, we impose the condition of pseudo-Hermiticity on $L^{}_j$, leading to
\begin{equation}
b^*_j - c^*_j = - (b^{}_j - c^{}_j) \,, \quad a^*_j - d^*_j = - (a^{}_j - d^{}_j) \,.
\end{equation}
The above constraints can be made more transparent if we adopt the following parametrization
\begin{equation}
L^{}_j = \left( \begin{matrix} r^{}_j e^{{\rm i}\varphi^{}_j} & s^{}_j e^{{\rm i}\phi^{}_j} \cr s^{}_j e^{-{\rm i}\phi^{}_j} & r^{}_j e^{-{\rm i} \varphi^{}_j}\end{matrix} \right) = r^{}_j \cos\varphi^{}_j \sigma^{}_0 + s^{}_j \cos\phi^{}_j \sigma^{}_1 - s^{}_j \sin\phi^{}_j \sigma^{}_2 + {\rm i} r^{}_j \sin\varphi^{}_j \sigma^{}_3 \,,
\end{equation}
where $r^{}_j$, $s^{}_j$, $\varphi^{}_j$, and $\phi^{}_j$ are all real parameters. It is worth mentioning that this parametrization has also been used for the non-Hermitian Hamiltonian in Eq.~(\ref{eq:H}), which is $\eta$-pseudo-Hermitian. Hence, the PT-symmetry is also conserved by the Lindblad operators, i.e.~$[L^{}_j, {\cal PT}] = \mathbb{0}^{}_2$, just like the Hamiltonian itself. Finally, using the definitions in Eqs.~(\ref{eq:A})--(\ref{eq:F}), we arrive at the explicit expressions for the Lindblad parameters as follows
\begin{align}
A &= \sum^3_{j=1} \left[(L^2_j)^2 + (L^3_j)^2\right] = \sum^3_{j=1} \left(s^2_j \sin^2 \phi^{}_j - r^2_j \sin^2 \varphi^{}_j \right) \,, \\
B &= \sum^3_{j=1} \left[(L^1_j)^2 + (L^3_j)^2\right] = \sum^3_{j=1} \left(s^2_j \cos^2 \phi^{}_j - r^2_j \sin^2 \varphi^{}_j \right) \,, \\
C &= \sum^3_{j=1} \left[(L^1_j)^2 + (L^2_j)^2\right] = \sum^3_{j=1} s^2_j \,, \\
D &= - \sum^3_{j=1} L^1_j L^2_j = \sum^3_{j=1} s^2_j \sin \phi^{}_j \cos \phi^{}_j \,, \\
E &= - \sum^3_{j=1} L^1_j L^3_j = - {\rm i} \sum^3_{j=1} r^{}_j s^{}_j \cos\phi^{}_j \sin\varphi^{}_j \,, \\
F &= - \sum^3_{j=1} L^2_j L^3_j = {\rm i} \sum^3_{j=1} r^{}_j s^{}_j \sin\phi^{}_j \sin\varphi^{}_j \,.
\end{align}
Two special cases are interesting and will be discussed in Sec.~\ref{sec:prob}. First, for $\phi^{}_j = \pi/2$ and $\varphi^{}_j = 0$ or $\pi$, we have $B = 0$, $A = C= \sum_{j=1}^3 s^2_j \geq 0$ and $D = E = F = 0$. Second, for $\phi^{}_j = \varphi^{}_j = 0$ or $\pi$, we obtain $A = 0$, $B = C = \sum_{j=1}^3 s^2_j \geq 0$ and $D = E = F = 0$. In both special cases, we are left with only one non-trivial parameter in the Lindblad term, which greatly simplifies the calculation of transition probabilities for between the eigenstates, as we will investigate next.

\section{Transition Probabilities}
\label{sec:prob}

\subsection{Strategy for Calculation of Transition Probabilities}

It is known that the density matrix $\rho$ can be expressed and parametrized in terms of the identity matrix $\sigma^{}_0 = \mathbb{1}^{}_2$ and the three Pauli matrices $\sigma^{}_i$ ($i = 1,2,3$) such that
\begin{equation}
\rho = \rho^{}_\mu \sigma^{}_\mu \equiv \rho^{}_0 \mathbb{1}^{}_2 + \rho^{}_i \sigma^{}_i = \left( \begin{matrix} \rho^{}_0 + \rho^{}_3 & \rho^{}_1 - {\rm i} \rho^{}_2 \cr \rho^{}_1 + {\rm i} \rho^{}_2 & \rho^{}_0 - \rho^{}_3 \end{matrix} \right) \,.
\end{equation}
The time evolution for the density matrix $\rho(t)$ with Hermitian and non-Hermitian Hamiltonians is given by the Lindblad equations in Eqs.~(\ref{eq:lindblad}) and (\ref{eq:Nlindblad}), respectively. A very clear presentation of the derivation of the Lindblad equation can be found in Ref.~\cite{Pearle:2012} and also in Ref.~\cite{Weinberg:2015}. The Lindblad equation for the generalized density matrix $\rho^{}_{\rm G}(t)$ has been presented in Refs.~\cite{Scolarici:2006, Scolarici:2007} and in Eq.~(\ref{eq:Glindblad}), and recast in a more constrained form in Eq.~(\ref{eq:Glindbladh}).

For the sake of self-consistency, we will implement the generalized density matrix $\rho^{}_{\rm G}(t)$ and its Lindblad equation in Eq.~(\ref{eq:Glindbladh}) to calculate the transition probabilities. The definition of a transition probability is the probability of transition from one of the two independent eigenstates of a given two-level quantum system to the other one. The two independent eigenstates of a two-level quantum system can be expressed in different eigenbases of the same Hilbert space, which is two-dimensional. In some two-level quantum systems, it is convenient to introduce two different eigenbases, which are related to each other by mixing, whereas in other systems, it is just useful to discuss one eigenbasis. For example, in the two-level quantum system describing neutrino oscillations, one normally introduces two eigenbases, i.e.~the mass eigenbasis and the flavor eigenbasis, where the former is constructed by the $+$ and $-$ ``mass'' states and the latter is described by the $a$ and $b$ ``flavor'' states to be discussed after Eq.~(\ref{eq:A-1}), which quantifies the mixing between the two eigenbases. The main procedure for our derivation of the transition probabilities is outlined in Appendix~\ref{app:procedure}. The first step is to specify the initial density matrices, which can be constructed from the normalized eigenvectors and the metric. Using Eqs.~(\ref{eq:neigenv}) and (\ref{eq:metric}), for the generalized density matrix $\rho^{}_{\rm G}(t)$, we obtain
\begin{align}
\rho^{}_{{\rm G},+}(0) &= |u^{}_+\rangle \langle u^{}_+| \eta = \frac{1}{2} \left( \begin{matrix}  1 + {\rm i} \tan \alpha & \sec \alpha \cr \sec \alpha &  1 - {\rm i} \tan \alpha  \end{matrix} \right) \,, \\
\rho^{}_{{\rm G},-}(0) &= |u^{}_-\rangle \langle u^{}_-| \eta = \frac{1}{2} \left( \begin{matrix} 1 - {\rm i} \tan \alpha &  -\sec \alpha \cr -\sec \alpha & 1 + {\rm i} \tan \alpha \end{matrix} \right) \,.
\end{align}
Note that it holds by construction that $\rho^{}_{{\rm G}, +}(0) + \rho^{}_{{\rm G}, -}(0) = \mathbb{1}^{}_2$ and ${\rm tr}\left[\rho^{}_{{\rm G}, +}(0)\right] = {\rm tr}\left[\rho^{}_{{\rm G}, -}(0)\right] = 1$. In the cases of $r = 0$ and $s = 0$, in the latter of which one needs to calculate the density matrix from the scratch by using Eq.~(\ref{eq:eigenvmetricr}), we find that the initial values for the generalized density matrices are
\begin{equation}
\rho^{r}_{{\rm G}, +}(0) = \left( \begin{matrix} 1 & 0 \cr 0 & 0 \end{matrix} \right) \,, \quad
\rho^{r}_{{\rm G}, -}(0) = \left( \begin{matrix} 0 & 0 \cr 0 & 1 \end{matrix} \right)
\end{equation}
and
\begin{equation}
\rho^{s}_{{\rm G}, +}(0) = \frac{1}{2} \left( \begin{matrix} 1 & 1 \cr 1 & 1 \end{matrix} \right) \,, \quad
\rho^{s}_{{\rm G}, -}(0) =  \frac{1}{2} \left( \begin{matrix} 1 & -1 \cr -1 & 1 \end{matrix} \right) \,,
\end{equation}
respectively. The transformation between the $2 \times 2$ density matrix $\rho$ and the corresponding vector $\Gamma$ with the four components of the original density matrix is given by
\begin{equation}
\rho = (\rho^{}_{ij}) = \left( \begin{matrix} \rho^{}_0 + \rho^{}_3 & \rho^{}_1 - {\rm i} \rho^{}_2 \cr \rho^{}_1 + {\rm i} \rho^{}_2 & \rho^{}_0 - \rho^{}_3 \end{matrix} \right) \quad \leftrightarrow \quad \Gamma = \left( \begin{matrix} \Gamma^{}_0 \cr \Gamma^{}_1 \cr \Gamma^{}_2 \cr \Gamma^{}_3 \end{matrix} \right) \,,
\label{eq:transf}
\end{equation}
where the relations between the components of $\rho$ and $\Gamma^{}_\mu$ (for $\mu = 0, 1, 2, 3$) are the following
\begin{align}
\Gamma^{}_0 &= \frac{\rho^{}_{11} + \rho^{}_{22}}{2} = \rho^{}_0 \,, \\
\Gamma^{}_1 &= \frac{\rho^{}_{12} + \rho^{}_{21}}{2} = \rho^{}_1 \,, \\
\Gamma^{}_2 &= \frac{\rho^{}_{21} - \rho^{}_{12}}{2 {\rm i}} = \rho^{}_2 \,, \\
\Gamma^{}_3 &= \frac{\rho^{}_{11} - \rho^{}_{22}}{2} = \rho^{}_3 \,.
\end{align}
Using Eq.~(\ref{eq:H}) and the evolution equations derived in Subsec.~\ref{sub:evoleq}, we obtain a Schr{\"o}dinger-like equation such that
\begin{equation}
\dot{\Gamma}(t) = -2 R \Gamma(t) \,,
\end{equation}
which has the solution
\begin{equation}
\Gamma(t) = \exp(-2 R t) \Gamma(0) \equiv {\cal M}(t) \Gamma(0) \,,
\end{equation}
where ${\cal M}(t)$ only depends on time $t$ and the eigenvalues of $R$, which should be the same in all bases. For the generalized density matrix, the matrix $R$, which is effectively a $4 \times 4$ ``Hamiltonian'', reads [see also Eq.~(\ref{eq:Glindblad3}) or Eq.~(\ref{eq:Glindblad4}) in Appendix~\ref{app:4Drep}]
\vspace{-1mm} 
\begin{equation}
R^{}_{\rm G} = \left( \begin{matrix} 0 & 0 & 0 & 0 \cr 0 & A & D + {\rm i} r \sin \varphi & E + s \sin \phi \cr 0 & D -{\rm i} r \sin \varphi & B & F + s \cos \phi \cr 0 & E -s \sin \phi & F -s \cos \phi & C \end{matrix} \right) \,.
\end{equation}
Under the assumption that $\phi = 0$ and the Lindblad parameter $A = 0$, which leads to $B = C$, $D = E = F = 0$ and means that there is only one non-zero Lindblad parameter $B = C = \xi$ (see e.g.~Ref.~\cite{Benatti:1997rv}), we obtain
\vspace{-1mm} 
\begin{equation}
R^{}_{\rm G}(\phi = 0, A = 0) = \left( \begin{matrix} 0 & 0 & 0 & 0 \cr 0 & 0 & {\rm i} r \sin \varphi & 0 \cr 0 & -{\rm i} r \sin \varphi & \xi & s \cr 0 & 0 & -s & \xi \end{matrix} \right) \,. \label{eq:RGA}
\end{equation}
In the other special case with $\phi = 0$ and the Lindblad parameter $B = 0$, we are left with $D = E = F = 0$ and only one non-zero Lindblad parameter $A = C = \zeta$. Thus, we find
\vspace{-1mm} 
\begin{equation}
R^{}_{\rm G}(\phi = 0, B = 0) = \left( \begin{matrix} 0 & 0 & 0 & 0 \cr 0 & \zeta & {\rm i} r \sin \varphi & 0 \cr 0 & -{\rm i} r \sin \varphi & 0 & s \cr 0 & 0 & -s & \zeta \end{matrix} \right) \,. \label{eq:RGB}
\end{equation}
Note that no simple analytic solutions for the eigenvalues of the most general form of $R^{}_{\rm G}(\phi = 0, A = 0)$ or $R^{}_{\rm G}(\phi = 0, B = 0)$ exist, and of course, also not for the even more general form of $R^{}_{\rm G}$.

The structure of $R$ is reflected in the structure of ${\cal M}(t) = \exp(-2 R t)$ due to the spectral decomposition theorem, which means that if $R$ relates some certain components of the density matrix, then ${\cal M}(t)$ will relate the same components, since the structure is not changed by the matrix exponentiation of $-2 R t$ that gives ${\cal M}(t)$. Thus, the solution to the Schr{\"o}dinger-like equation $\Gamma(t) = {\cal M}(t) \Gamma(0)$, which is the time evolution for the vector of the density matrix components (with the initial density matrix components as the initial condition), will lead to the time evolution for the density matrix $\rho(t)$ itself and is naturally encoded in ${\cal M}(t)$. Therefore, using Eq.~(\ref{eq:transf}), we can transform the vector of the density matrix components $\Gamma(t)$ back to the $2 \times 2$ density matrix $\rho(t)$. Thus, under the assumption that $\phi = 0$ and $A = 0$, we obtain (including the initial condition given by the initial density matrix under consideration)
\begin{equation}
\rho^{}_{\rm G}(t) = \left( \begin{matrix} \Gamma^{}_{{\rm G},0}(0) + \Gamma^{}_{{\rm G},3}(t) & \Gamma^{}_{{\rm G},1}(t) - {\rm i} \Gamma^{}_{{\rm G},2}(t)  \cr \Gamma^{}_{{\rm G},1}(t) + {\rm i} \Gamma^{}_{{\rm G},2}(t) & \Gamma^{}_{{\rm G},0}(0) - \Gamma^{}_{{\rm G},3}(t) \end{matrix} \right) \,,
\end{equation}
where $\Gamma^{}_{{\rm G},0}(t) = \Gamma^{}_{{\rm G},0}(0) = \mbox{const.}$ due to the structures of $R^{}_{\rm G}(\phi = 0, A = 0)$ in Eq.~(\ref{eq:RGA}) and $R^{}_{\rm G}(\phi = 0, B = 0)$ in Eq.~(\ref{eq:RGB}), respectively. Note that the trace ${\rm tr} \left[\rho^{}_{\rm G}(t) \right] = 2 \Gamma^{}_{{\rm G},0}(0) = 2 \rho^{}_{{\rm G},0}(0)$ is time independent. Now, the transition probability between the $|u^{}_+\rangle$ and $|u^{}_-\rangle$ states is given by
\begin{equation}
P_{+-}^{\rm G}(t) = {\rm tr}\left[\rho^{}_{{\rm G},+}(t) \rho^{}_{{\rm G},-}(0)\right]
\end{equation}
for the generalized density matrix, where $\rho^{}_{{\rm G},+}(t)$ is the solution of $\rho^{}_{\rm G}(t)$ with the initial condition $\rho^{}_{\rm G}(0) = \rho^{}_{{\rm G},+}(0)$. In a similar way, one can compute the other transition probabilities $P^{\rm G}_{++}(t)$, $P^{\rm G}_{-+}(t)$, and $P^{\rm G}_{--}(t)$.

Using the definition for the mixing matrix $A^{}_{\rm inv}$ in Ref.~\cite{Ohlsson:2019noy} (based on the normalized $|u^{}_+\rangle$ and $|u^{}_-\rangle$ eigenvectors), we can derive
\begin{equation}
{\cal A}^{-1} \equiv A_{\rm inv} = \frac{1}{\sqrt{2\cos\alpha}} \left( \begin{matrix} e^{{\rm i} \alpha/2} & e^{-{\rm i} \alpha/2} \cr e^{-{\rm i} \alpha/2} & e^{{\rm i} \alpha/2} \end{matrix} \right) \,.
\label{eq:A-1}
\end{equation}
Transforming from the $+$ and $-$ ``mass'' states to the $a$ and $b$ ``flavor'' states by applying the mixing matrix ${\cal A}^{-1}$, we obtain
\begin{align}
|u^{}_a\rangle &= \left({\cal A}^{-1}\right)^{}_{a+} |u^{}_+\rangle + \left({\cal A}^{-1}\right)^{}_{a-} |u^{}_-\rangle = \left( \begin{matrix} 1 & 0 \end{matrix} \right)^{\rm T} \,, \label{eq:ua}\\
|u^{}_b\rangle &= \left({\cal A}^{-1}\right)^{}_{b+} |u^{}_+\rangle + \left({\cal A}^{-1}\right)^{}_{b-} |u^{}_-\rangle = \left( \begin{matrix} 0 & 1 \end{matrix} \right)^{\rm T} \,. \label{eq:ub}
\end{align}
Thus, the $a$ and $b$ ``flavor'' states are linear superpositions of the $+$ and $-$ ``mass'' states, where the different elements of the mixing matrix ${\cal A}^{-1}$ in Eq.~(\ref{eq:A-1}) are the coefficients in the two superpositions given by Eqs.~(\ref{eq:ua}) and (\ref{eq:ub}), respectively. As mentioned, for some two-level quantum systems, like neutrino oscillations, it is convenient to use two different sets of states, i.e.~the $a$ and $b$ ``flavor'' states and the $+$ and $-$ ``mass'' states, whereas for other systems, these two sets of states coincide or it is only useful to discuss one of the two sets. In the situation that only one set of states is necessary to describe physics, we will refer to the two states in that set of states (the eigenbasis) as the $a$ and $b$ states, i.e.~the physical states. It is interesting to note that for both the general case ($r \neq 0$ and $s \neq 0$) and the cases when $r = 0$ or $s = 0$, the $a$ and $b$ ``flavor'' states will be the same and given by Eqs.~(\ref{eq:ua}) and (\ref{eq:ub}). The initial ``flavor'' density matrices can then be constructed using the $a$ and $b$ ``flavor'' states and the metric. For the generalized density matrix, we find that
\begin{equation}
\rho^{}_{{\rm G},a}(0) = |u^{}_a\rangle \langle u^{}_a| \eta = \left( \begin{matrix} \sec \alpha & -{\rm i} \tan \alpha \cr 0 & 0 \end{matrix} \right) \,, \quad
\rho^{}_{{\rm G},b}(0) = |u^{}_b\rangle \langle u^{}_b| \eta = \left( \begin{matrix} 0 & 0 \cr {\rm i} \tan \alpha & \sec \alpha \end{matrix} \right) \,,
\label{eq:rhoGab}
\end{equation}
where ${\rm tr}\left[\rho^{}_{{\rm G},a}(0)\right] = {\rm tr} \left[\rho^{}_{{\rm G},b}(0)\right] = \sec \alpha$ holds. For the cases $r = 0$ and $s = 0$, we have
\begin{equation}
\rho_{{\rm G},a}^{r,s}(0) = \left( \begin{matrix} 1 & 0 \cr 0 & 0 \end{matrix} \right) \,, \quad
\rho_{{\rm G},b}^{r,s}(0) = \left( \begin{matrix} 0 & 0 \cr 0 & 1 \end{matrix} \right) \,,
\end{equation}
which can be obtained from Eq.~(\ref{eq:rhoGab}) by setting $\alpha = 0$. It should be noted that $\rho_{{\rm G},\pm}^{r}(0)$ and $\rho_{{\rm G},a(b)}^{r}(0)$ are the same, since the ``mass'' state basis and the ``flavor'' state basis coincide in this case. The reason is that the mixing matrix ${\cal A}^{-1}$ is trivial, which follows from the fact that the Hamiltonian is diagonal (although complex), and therefore, there is basically only one interesting basis to consider. However, in order to have mixing, one can introduce an arbitrary mixing characterized by an angle $\theta$ (see Ref.~\cite{Ohlsson:2000mj}), which means that the initial density matrices would be
\begin{equation}
\rho^{}_{\alpha}(0) =  \left( \begin{matrix} \cos^2 \theta & \sin \theta \cos \theta \\ \sin \theta \cos \theta & \sin^2 \theta \end{matrix} \right) \,, \quad \rho^{}_{\beta}(0) =  \left( \begin{matrix} \sin^2 \theta & -\sin \theta \cos \theta \\ -\sin \theta \cos \theta & \cos^2 \theta \end{matrix} \right) \,,
\end{equation}
where ${\rm tr}\left[\rho^{}_\alpha(0)\right] = {\rm tr} \left[\rho^{}_\beta(0)\right] = \cos^2 \theta + \sin^2 \theta = 1$ is normalized automatically.

Concerning the initial generalized density matrices in Eq.~(\ref{eq:rhoGab}), they are, after the rotation to the $a$ and $b$ ``flavor'' basis with the mixing matrix ${\cal A}^{-1}$, no longer normalized (i.e.~${\rm tr} \left[ \rho^{}_{{\rm G},a}(0) \right] = {\rm tr} \left[ \rho^{}_{{\rm G},b}(0) \right] = \sec \alpha \neq 1$). Therefore, one must normalize them (since the completeness relation changes due to the metric being non-trivial) in order to obtain transition probabilities that are restricted between 0 and 1. The initial normalized generalized density matrices are given by
\begin{equation}
\rho^{}_{{\rm GN},a}(0) = \frac{\rho^{}_{{\rm G},a}(0)}{{\rm tr}\left[ \rho^{}_{{\rm G},a}(0) \right]} = \left( \begin{matrix} 1 & -{\rm i} \sin \alpha \cr 0 & 0 \end{matrix} \right) \,, \quad
\rho^{}_{{\rm GN},b}(0) = \frac{\rho^{}_{{\rm G},b}(0)}{{\rm tr} \left[ \rho^{}_{{\rm G},b}(0) \right]} = \left( \begin{matrix} 0 & 0 \cr {\rm i} \sin \alpha & 1 \end{matrix} \right) \,, \label{eq:rhoGNab}
\end{equation}
which indeed have ${\rm tr} \left[\rho^{}_{{\rm GN},a}(0) \right] = {\rm tr} \left[\rho^{}_{{\rm GN},b}(0) \right] = 1$. Finally, the transition probability from the ``flavor'' state $a$ to the ``flavor'' state $b$ is given by
\begin{equation}
P_{ab}^{\rm G}(t) = {\rm tr} \left[ \rho^{}_{{\rm G},a}(t) \rho^{}_{{\rm G},b}(0) \right] \qquad {\rm or} \qquad P_{ab}^{\rm G}(t) = {\rm tr} \left[ \rho^{}_{{\rm G},a}(t) \rho^{}_{{\rm GN},b}(0) \right] \,,
\end{equation}
where the initial density matrices in the latter case are normalized. Note that in what follows, except in connection to neutrino oscillations and in Appendix~\ref{app:procedure}, we will refer to the $+$ and $-$ ``mass'' states and the $a$ and $b$ ``flavor'' states as just the $+$ and $-$ states and $a$ and $b$ states, respectively.

\subsection{The Lindblad Term with $\mbox{\boldmath$A = 0$}$}
\label{sec:LindbladA}

\subsubsection{General Discussion}

Due to the fact that the characteristic equation of $R^{}_{\rm G}(\phi = 0, A = 0)$, which can be reduced to a $3 \times 3$ matrix that we simply call $R$, is a general cubic algebraic equation on the form
\begin{equation}
\lambda^3 + c^{}_2 \lambda^2 + c^{}_1 \lambda + c^{}_0 = 0 \,,
\label{eq:cubic_general}
\end{equation}
where the coefficients $c^{}_n$ ($n = 0,1,2)$ are given by the principal invariants
$$
c^{}_2 = - {\rm tr \,} R \,, \quad c^{}_1 = \frac{1}{2} \left[ ({\rm tr \,} R)^2 - {\rm tr \,} (R^2) \right] \,, \quad c^{}_0 = - \det R \,.
$$
The roots of the characteristic equation in Eq.~(\ref{eq:cubic_general}), i.e.~the eigenvalues, will not be simple closed-form expressions in terms of the parameters $r$, $s$, $\varphi$, and $\xi$. Therefore, we will first investigate approximate formulas for the transition probabilities, and second, some special cases (with specific choices of the parameters). Despite the fact that there are no simple closed-form expressions of the eigenvalues, the general form of the cubic characteristic equations given in Eq.~(\ref{eq:cubic_general}) can be heavily reduced to so-called depressed cubic characteristic equations on the form
\begin{equation}
\lambda^3 + c^{}_1 \lambda + c^{}_0 = 0
\label{eq:depressed}
\end{equation}
by transforming $R$ to another matrix
\begin{equation}
S \equiv R - \frac{{\rm tr \,} R}{3} \mathbb{1}_3 \,,
\end{equation}
which leads to $c_2 = - {\rm tr \,} S = 0$. The roots of the depressed cubic characteristic equations~(\ref{eq:depressed}) are simpler expressions than those of the general cubic characteristic equations~(\ref{eq:cubic_general}), and thus easier to handle in order to obtain some useful results. By employing the method described in Ref.~\cite{Ohlsson:1999xb}, sometimes referred to as the Ohlsson--Snellman method that is based on the Cayley--Hamilton theorem, we can express the matrix exponential of $R$ exactly as
\begin{equation}
\exp(-2 R t) = \Phi \left( \mathbb{1}_3 a^{}_0 - 2 S t a^{}_1 + 4 S^2 t^2 a^{}_2 \right) \,, \quad \Phi = \exp\left(- 2 t \, {\rm tr \,} R/3 \right) \,,
\label{eq:expR}
\end{equation}
where the coefficients $a^{}_n$ ($n = 0,1,2$) can readily be determined from the following linear system of equations [in terms of the three roots of the characteristic equation $\lambda_i $ ($i = 1,2,3$) for $S$ on the form~(\ref{eq:depressed})], viz.,
\begin{align}
\exp(-2 \lambda_1 t) &= a^{}_0 - 2 \lambda^{}_1 t a^{}_1 + 4 \lambda_1^2 t^2 a^{}_2 \,, \label{eq:linear1}\\
\exp(-2 \lambda_2 t) &= a^{}_0 - 2 \lambda^{}_2 t a^{}_1 + 4 \lambda_2^2 t^2 a^{}_2 \,, \label{eq:linear2}\\
\exp(-2 \lambda_3 t) &= a^{}_0 - 2 \lambda^{}_3 t a^{}_1 + 4 \lambda_3^2 t^2 a^{}_2 \,.
\label{eq:linear3}
\end{align}
Note that this method is exact for computing $\exp(-2 R t)$ in Eq.~(\ref{eq:expR}) and does not rely on any approximations. In fact, the series $\exp(-2 R t) = \sum_{n=0}^\infty (-2 R t)^n/{n!}$ is cut after three terms due to the Cayley--Hamilton theorem \cite{Ohlsson:1999xb}. Consequently, solving Eq.~(\ref{eq:depressed}) for the three eigenvalues of $S$, inserting them into the linear system of equations in Eqs.~(\ref{eq:linear1})--(\ref{eq:linear3}) in order to solve for $a^{}_n$ ($n = 0,1,2$) which in turn are inserted into Eq.~(\ref{eq:expR}), we exactly obtain ${\cal M}(t) = \exp(-2 R t)$, which leads to the time evolution $\Gamma(t) = {\cal M}(t) \Gamma(0)$. Despite the method being exact, it will not lead to expressions for the transition probabilities that are simple. However, the exact expressions can now be used to derive approximate expressions for the transition probabilities. Of course, the exact expressions can also be used to compute the transition probabilities numerically.

\begin{figure}[t!]
\begin{center}
\begin{tabular}{cc}
\includegraphics[width=0.48\textwidth]{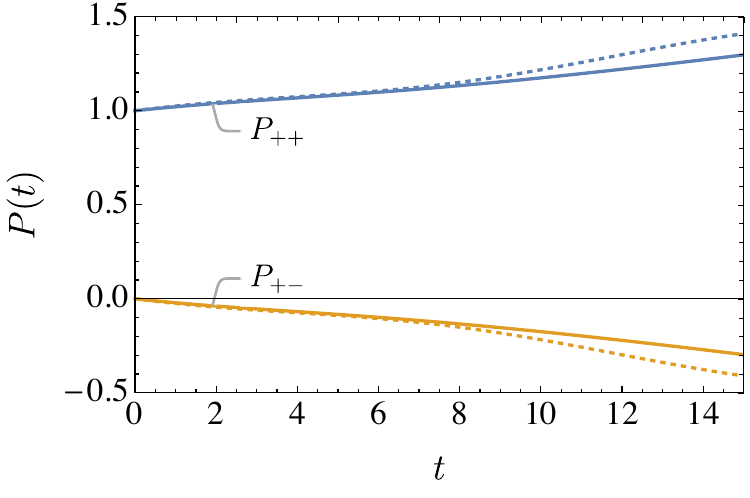} & \includegraphics[width=0.48\textwidth]{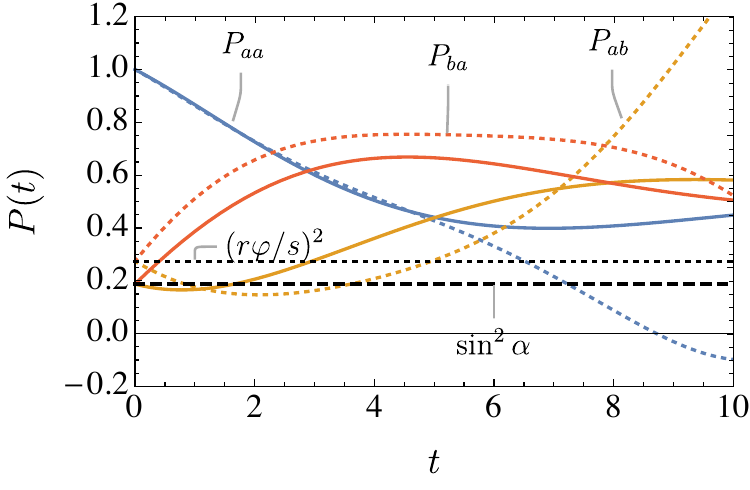}\\
\end{tabular}
\end{center}
\vspace{-5mm}
\caption{Exact numerical (solid curves) and approximate analytical (dotted curves) transition probabilities as functions of time $t$. {\it Left:} Transition probabilities for $+$ and $-$ states. {\it Right:} Transition probabilities for $a$ and $b$ states. Parameter values: $r = 0.1$, $s = 0.2$, $\varphi = \pi/3$, $\xi = 0.1$.}
\label{fig:approx}
\end{figure}

Thus, using the described method for Eq.~(\ref{eq:RGA}) and assuming that $\varphi$ and $\xi$ are small para\-meters, we series expand up to second order in $\varphi$ and $\xi$  and obtain approximate analytical formulas for the transition probabilities between the $+$ and $-$ states as
\begin{align}
P_{++}(t) = P_{--}(t) &= 1 + \frac{r^2 t}{s^2} \varphi^2 \xi - \frac{r^2 \sin^2(s t)}{s^4} \varphi^2 \xi^2 + {\cal O}(\varphi^3,\xi^3) \,, \label{eq:P++approx} \\
P_{+-}(t) = P_{-+}(t) &= - \frac{r^2 t}{s^2} \varphi^2 \xi + \frac{r^2 \sin^2(s t)}{s^4} \varphi^2 \xi^2 + {\cal O}(\varphi^3,\xi^3) \,, \label{eq:P+-approx}
\end{align}
whose numerical results are shown as functions of time $t$  and plotted as dotted curves in the left panel of Fig.~\ref{fig:approx}. It is important to note that $\varphi$ and $\xi$ are not small parameters in all relevant physical situations. Thus, the validity of this approximation only holds for two-level quantum systems in which it can be argued that (i) dissipative effects, described by $\xi$, are small and (ii) the diagonal elements of the Hamiltonian in Eq.~(\ref{eq:H}) are nearly equal, quantified by $\varphi$. For a two-level quantum system that fulfill these two criteria, the approximation is certainly valid. In our numerical calculations, the parameter values $r = 0.1$, $s = 0.2$, $\varphi = \pi/3$, $\xi = 0.1$ are chosen, i.e.~$\xi$ is indeed small and $\varphi$ is not large. One can observe that the approximate analytical results are well consistent with the exact numerical ones (see solid curves) for $t \lesssim 9$, but significant deviations show up for $t \gtrsim 9$. This can be ascribed to the secular factors appearing in the series expansions, such as $\xi t$, which will no longer be small for large values of $t$. In addition, $P^{}_{++}(t) = P^{}_{--}(t)$ and they are larger than one and increase with $t$, while $P^{}_{+-}(t) = P^{}_{-+}(t)$ and they turn out to be negative. However, $P^{}_{+-}(t) = P^{}_{-+}(t)$ vanishes in the limit of $\xi \to 0$, implying the absence of the Lindblad term. This should be the case, since there are no transitions between the $+$ and $-$ states for the PT-symmetric Hamiltonian. Furthermore, one can observe that the transition probabilities between the $+$ and $-$ states sum up to unity, i.e.~$P_{++}(t) + P_{+-}(t) = P_{++}(t) + P_{-+}(t) = P_{--}(t) + P_{-+}(t) = P_{--}(t) + P_{+-}(t) = 1 + {\cal O}(\varphi^3, \xi^3)$, which they must in order for conservation of probability to hold. The term $(r^2 \varphi^2/s^2) \xi t$ [and also the term $(r^2 \sin^2(st)/s^4) \varphi^2 \xi^2 \simeq (r^2 \varphi^2/s^2) (\xi t)^2$] is basically a measure of dissipative effects between the $+$ and $-$ states and also a consequence of the non-trivial metric used. Thus, in the presence of dissipative effects, the individual transition probabilities between the $+$ and $-$ states in Eqs.~(\ref{eq:P++approx}) and (\ref{eq:P+-approx}) are not restricted to values between 0 and 1, and therefore, it is not possible to consider a given state and determine its transition probabilities to itself or to the other state, whereas this is true for the conservation of probability, which is physical, i.e.~there is no probability disappearing from the total two-level system (read: both of the two states). However, if the dissipative effects are turned off, i.e.~$\xi \to 0$, then $P_{++}(t) = P_{--}(t) = 1$ and $P_{+-}(t) = P_{-+}(t) = 0$, and the individual transition probabilities between the $+$ and $-$ states will be restricted to either 0 or 1.


For the transition probabilities between the $a$ and $b$ states, we can also derive approximate analytical formulas and obtain
\begin{align}
P_{aa(bb)}(t) &= \cos^2(st) - t \cos(2st) \xi + t^2 \cos(2st) \xi^2 + \frac{r^2 t \sin(2st)}{2s} \varphi^2 \nonumber\\
&- \frac{r^2 \{\sin(2st) + 2 st [\cos(2st) + 2st \sin(2st)]\}}{4s^3} \varphi^2 \xi + \frac{r^2 t^2 [\cos(2st) + st \sin(2st)]}{s^2} \varphi^2 \xi^2 \nonumber\\
&+ {\cal O}(\varphi^3,\xi^3) \,, \label{eq:paaapprox}\\
P_{ab,ba}(t) &= \sin^2(st) \mp \frac{r \sin(2st)}{s} \varphi + t \cos(2st) \xi \pm \frac{2rt \sin(2st)}{s} \varphi \xi + \frac{r^2 [2 \cos(2st) - st \sin(2st)]}{2 s^2} \varphi^2 \nonumber\\
&- t^2 \cos(2st) \xi^2 \mp \frac{2 rt^2 \sin(2st)}{s} \varphi \xi^2 + \frac{r^2 [(1+4s^2t^2)\sin(2st)-6st \cos(2st)]}{4 s^3} \varphi^2 \xi \nonumber\\
&+ \frac{r^2 t^2 [\cos(2st)-st\sin(2st)]}{s^2} \varphi^2 \xi^2 + {\cal O}(\varphi^3,\xi^3) \,, \label{eq:pabapprox}
\end{align}
where $P^{}_{aa}(t) = P^{}_{bb}(t)$ is found in Eq.~(\ref{eq:paaapprox}) and the upper and lower signs in Eq.~(\ref{eq:pabapprox}) refer to $P^{}_{ab}(t)$ and $P^{}_{ba}(t)$, respectively. Both approximate analytical (dotted curves) and exact numerical (solid curves) results of these transition probabilities are presented in the right panel of Fig.~\ref{fig:approx}. Three important observations can be made. First, at the initial time $t = 0$, one can see from Eq.~(\ref{eq:paaapprox}) that $P^{}_{aa}(0) = P^{}_{bb}(0) = 1$, which is exactly the same as the numerical result. For $t = 0$, the initial generalized density matrices have been given in Eq.~(\ref{eq:rhoGab}), and thus, the survival probabilities can be calculated as $P^{}_{aa}(0) = {\rm tr}\left[\rho^{}_{{\rm GN}, a}(0)\rho^{}_{{\rm GN}, a}(0)\right] = 1$ and $P^{}_{bb}(0) = {\rm tr}\left[\rho^{}_{{\rm GN}, b}(0)\rho^{}_{{\rm GN}, b}(0)\right] = 1$. For $t \lesssim 5$, there is an excellent agreement between analytical and numerical results of $P^{}_{aa}(t) = P^{}_{bb}(t)$. Second, from Eq.~(\ref{eq:pabapprox}), we can find $P^{}_{ab}(0) = P^{}_{ba}(0) = (r\varphi/s)^2$, which is displayed as the dotted horizontal line in the right panel of Fig.~\ref{fig:approx}. However, comparing the solid and dotted curves for $P_{ab}(t) = P_{ba}(t)$, one can notice that there is a sizable discrepancy between exact numerical and approximate analytical results, even at the initial time $t = 0$. With the help of Eq.~(\ref{eq:rhoGNab}), we can obtain  $P^{}_{ab}(0) = {\rm tr}\left[\rho^{}_{{\rm GN}, a}(0)\rho^{}_{{\rm GN}, b}(0)\right] = \sin^2\alpha$ and $P^{}_{ba}(0) = {\rm tr}\left[\rho^{}_{{\rm GN}, b}(0)\rho^{}_{{\rm GN}, a}(0)\right] = \sin^2\alpha$, which are shown as the dashed horizontal line in the right panel of Fig.~\ref{fig:approx}. Note that $\sin^2\alpha = (r\sin\varphi/s)^2$ approaches $(r\varphi/s)^2$ only if $\varphi$ is small, which is not the situation in Fig.~\ref{fig:approx}. Third, and most importantly, in contrast to the individual transition probabilities between the $+$ and $-$ states, the individual transition probabilities between the $a$ and $b$ states are always restricted to values between 0 and 1 using the exact numerical results and when the approximate analytical results hold. However, on the other hand, conservation of probability does not hold for the transition probabilities between the $a$ and $b$ states at all times $t$, which is an effect of the non-trivial metric ($\sin^2 \alpha = r^2 \sin^2 \varphi/s^2 \neq 0$). Using Eqs.~(\ref{eq:paaapprox}) and (\ref{eq:pabapprox}), we find that
\begin{align}
P_{aa}(t) + P_{ab}(t) &= 1 + \frac{r \varphi}{s^2} \left[r \varphi \cos(2 st) - s \sin (2 st)\right] \left[1 + 2 \xi t (\xi t - 1)\right] + {\cal O}(\varphi^3,\xi^3) \,, \label{eq:PaaPab} \\
P_{bb}(t) + P_{ba}(t) &= 1 + \frac{r \varphi}{s^2} \left[r \varphi \cos(2 st) + s \sin (2 st)\right] \left[1 + 2 \xi t (\xi t - 1)\right] + {\cal O}(\varphi^3,\xi^3) \,. \label{eq:PbbPba}
\end{align}
In conclusion, for the $+$ and $-$ states, conservation of probability holds for the transition probabilities, whereas for the $a$ and $b$ states, the individual transition probabilities are restricted to values between 0 and 1. Thus, the mixing between the two different eigenbases determines which of the two bases that leads to a proper physical description of the transition probabilities of the two-level system in a given situation. If the mixing is trivial, i.e.~$\alpha \to 0$, then the two eigenbases coincide and there is only one physical basis (and only one set of physical states). Finally, in order to solve the problem of probability non-conservation for the transition probabilities between the $a$ and $b$ states, one may renormalize the individual transition probabilities for the $a$ and $b$ states and remove the effect of the non-trivial metric in a proper way, as advocated in Ref.~\cite{Ohlsson:2019noy}.

\subsubsection{Specific Examples with $\mbox{\boldmath$A = 0$}$}

After presenting the approximate analytical formulas for the transition probabilities in the general case, we investigate some specific examples with both analytical and numerical results, and explain the main features of the dissipative effects induced by the Lindblad term.
\begin{enumerate}
\item \label{case:2fneuosc} {\it Two-flavor neutrino oscillations.} This is a simple example of a Hermitian Hamiltonian $H = H^\dagger$ (with a trivial metric $\eta = \mathbb{1}^{}_2$) that has been extensively discussed in the literature~\cite{Benatti:2000ph}. In this case, the Hamiltonian $H$ and the Lindblad equation can be written as
\begin{equation}
H = \left( \begin{matrix} -\omega & 0 \cr 0 & \omega \end{matrix} \right) \quad \Leftrightarrow \quad \left( \begin{matrix} \dot{\rho}_0 \cr \dot{\rho}_1 \cr \dot{\rho}_2 \cr \dot{\rho}_3 \end{matrix} \right) = -2 \underbrace{\left( \begin{matrix} 0 & 0 & 0 & 0 \cr 0 & 0 & -\omega & 0 \cr 0 & \omega & \xi & 0 \cr 0 & 0 & 0 & \xi \end{matrix} \right)}_R \left( \begin{matrix} \rho_0 \cr \rho_1 \cr \rho_2 \cr \rho_3 \end{matrix} \right) \,,
\end{equation}
implying that $\rho^{}_0(t) = \rho_0$ is constant. Note that $H$ has been given in ``mass'' basis with $\omega \equiv (m^2_2 - m^2_1)/(4E)$, where $m^{}_1$ and $m^{}_2$ are the two neutrino masses and $E$ is the neutrino energy. The structure of $R$ relates components $\rho^{}_1$, $\rho^{}_2$, and $\rho^{}_3$ of the density matrix $\rho$ and the time evolution is governed by $H \rho - \rho H$, since $H^\dagger = H$. As a consequence, the trace of the density matrix is equal to $2 \rho_0(0) = 1$, which means that $\rho_0(0) = 1/2$. We find the transition probabilities for the ``mass'' and ``flavor'' states as
\begin{align}
P_{++} = P_{--} &= \frac{1}{2} \left( 1 + e^{-2 \xi t} \right) \,, \\
P_{+-} = P_{-+} &= \frac{1}{2} \left( 1 - e^{-2 \xi t} \right) \,, \\
P_{aa} = P_{bb} &= \frac{1}{2} \left[ 1 + e^{-2 \xi t} \cos^22\theta + e^{-\xi t} \sin^22\theta \left( \cos \Omega t + \frac{\xi}{\Omega} \sin \Omega t \right) \right] \,, \\
P_{ab} = P_{ba} &= \frac{1}{2} \left[ 1 - e^{-2 \xi t} \cos^22\theta - e^{-\xi t} \sin^22\theta \left( \cos \Omega t + \frac{\xi}{\Omega} \sin \Omega t \right) \right] \,,
\end{align}
where $\Omega \equiv \sqrt{4 \omega^2 - \xi^2}$. The formulas for $P^{}_{aa} = P^{}_{bb}$ and $P^{}_{ab} = P^{}_{ba}$ should be compared to the ones in Ref.~\cite{Benatti:2000ph}. In the limit $\xi \to 0$ (i.e.~without dissipative effects), we have $P_{++} = P_{--} = 1$ and $P_{+-} = P_{-+} = 0$, whereas $P_{aa} = P_{bb} = 1 - \sin^2 2\theta \sin^2 \omega t$ and $P_{ab} = P_{ba} = \sin^2 2\theta \sin^2 \omega t$, which are the ordinary formulas for two-flavor neutrino oscillations. Furthermore, for large values of $t$, all transition probabilities will average to 1/2. In Fig.~\ref{fig:2flneuosc}, we show the formulas for the two-flavor neutrino oscillation case with a specific choice for the parameter values ($\omega = 0.2$, $\xi = 0.1$, and $\theta = \pi/3$), which are not necessarily realistic and close to values supported by neutrino experiments. We note that both the transition probabilities for the ``mass'' and ``flavor'' states are damped by $\xi$.
\begin{figure}[t!]
\begin{center}
\begin{tabular}{cc}
\includegraphics[width=0.48\textwidth]{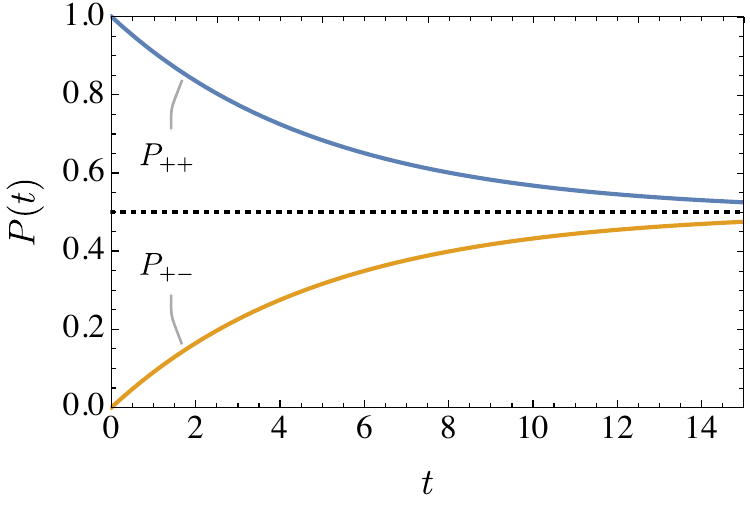} & \includegraphics[width=0.48\textwidth]{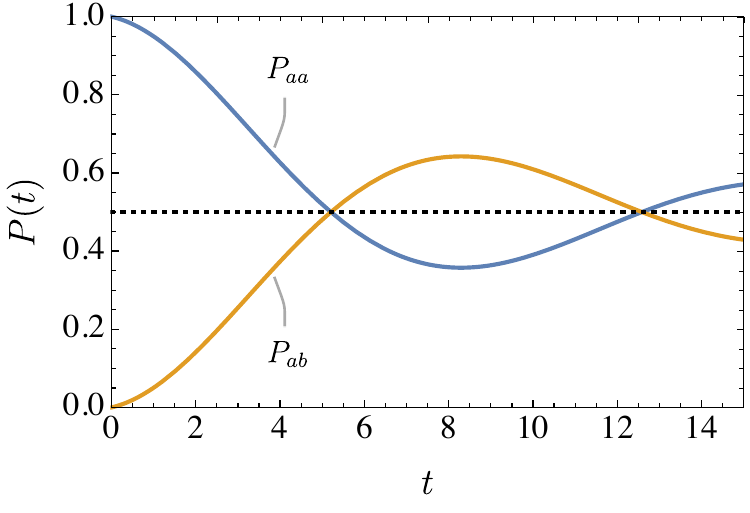}\\
\end{tabular}
\end{center}
\vspace{-5mm}
\caption{Two-flavor neutrino oscillations. {\it Left:} Transition probabilities for ``mass'' states. {\it Right:} Transition probabilities for ``flavor'' states. Parameter values: $\omega = 0.2$, $\xi = 0.1$, $\Omega \approx 0.39$, $\theta = \pi/3$.}
\label{fig:2flneuosc}
\end{figure}

\item \label{case:PTwoLindblad} {\it PT-symmetric non-Hermitian Hamiltonian without Lindblad term}. This case has been discussed in Ref.~\cite{Ohlsson:2019noy} without implementing the density matrix formalism.  In this case, the Hamiltonian $H$ is non-Hermitian and the metric $\eta$ is then non-trivial. The Hamiltonian and the ``Lindblad equation'' (actually the Liouville--von Neumann equation) for the generalized density matrix $\rho^{}_{\rm G}$ read
\begin{equation}
H = \left( \begin{matrix} r e^{{\rm i} \varphi} & s \cr s & r e^{-{\rm i} \varphi} \end{matrix} \right) \quad \Leftrightarrow \quad \left( \begin{matrix} \dot{\rho}^{}_{{\rm G},0} \cr \dot{\rho}^{}_{{\rm G},1} \cr \dot{\rho}^{}_{{\rm G},2} \cr \dot{\rho}^{}_{{\rm G},3} \end{matrix} \right) = -2 \underbrace{\left( \begin{matrix} 0 & 0 & 0 & 0 \cr 0 & 0 & {\rm i} r \sin \varphi & 0 \cr 0 & -{\rm i} r \sin \varphi & 0 & s \cr 0 & 0 & -s & 0 \end{matrix} \right)}_{R^{}_{\rm G}} \left( \begin{matrix} \rho^{}_{{\rm G},0} \cr \rho^{}_{{\rm G},1} \cr \rho^{}_{{\rm G},2} \cr \rho^{}_{{\rm G},3} \end{matrix} \right) \,,
\end{equation}
where $H$ is exactly the same as in Eq.~(\ref{eq:H}) with $\phi = 0$. Since the metric is non-trivial, $\rho^{}_{\rm G}$ should be used and the time evolution is governed by $H \rho^{}_{\rm G} - \rho^{}_{\rm G} H$. The structure of $R^{}_{\rm G}$ relates components $\rho^{}_{{\rm G},1}$, $\rho^{}_{{\rm G},2}$, and $\rho^{}_{{\rm G},3}$ of the generalized density matrix, and thus, the trace of the generalized density matrix is equal to $2 \rho^{}_{{\rm G},0}(0) = 1$, which means that $\rho^{}_{{\rm G},0}(0) = 1/2$. For the transition probabilities between the $a$ and $b$ states, the present case reproduces Eqs.~(2.24)--(2.27) in Ref.~\cite{Ohlsson:2019noy}. However, in order for these transition probabilities to be restricted between 0 and 1, the normalized generalized density matrix must be used after mixing with $A_{\rm inv}$, since the completeness relation changes due to the metric being non-trivial. Thus, we obtain the transition probabilities for the $+$ and $-$ states and the $a$ and $b$ states as
\begin{align}
P_{++} = P_{--} &= 1 \,, \\
P_{+-} = P_{-+} &= 0 \,, \\
P_{aa} = P_{bb} &= \cos^2 \frac{\beta t}{2} \,, \\
P_{ab} &= \sin^2 \left( \alpha - \frac{\beta t}{2} \right) \,, \\
P_{ba} &= \sin^2 \left( \alpha + \frac{\beta t}{2} \right) \,,
\end{align}
where $\alpha \equiv \arcsin(r \sin \varphi/s)$ and $\beta \equiv 2 \sqrt{s^2 - r^2 \sin^2 \varphi}$ are defined as in Ref.~\cite{Ohlsson:2019noy}. It should be noted that $P_{aa}(t=0) = P_{bb}(t=0) = 1$, whereas $P_{ab}(t=0) = P_{ba}(t=0) = \sin^2 \alpha \neq 0$, which is an effect of the non-trivial metric. In Fig.~\ref{fig:PTwoLindblad}, we present the transition probabilities for the $a$ and $b$ states, since the ones for the $+$ and $-$ states are trivial. We have exactly reproduced the results in Ref.~\cite{Ohlsson:2019noy}, when they are properly normalized. As one can observe from Fig.~\ref{fig:PTwoLindblad}, the transition probabilities oscillate with respect to time $t$, and no dissipative effects are found due to the absence of Lindblad operators.
\begin{figure}[t!]
\begin{center}
\includegraphics[width=0.5\textwidth]{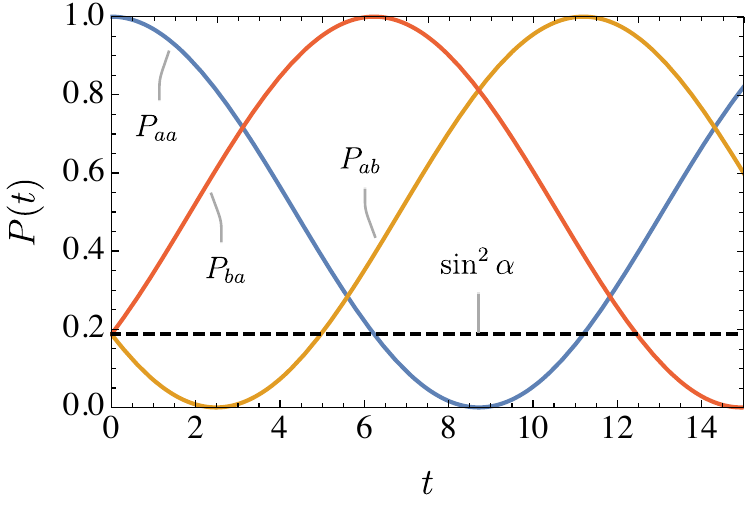}
\end{center}
\vspace{-5mm}
\caption{Transition probabilities for $a$ and $b$ states in the case of a PT-symmetric non-Hermitian Hamiltonian without Lindblad term. Parameter values: $r = 0.1$, $s = 0.2$, $\varphi = \pi/3$, $\alpha \approx 0.45$, $\beta \approx 0.36$.}
\label{fig:PTwoLindblad}
\end{figure}

\item \label{case:PTs=0} {\it PT-symmetric non-Hermitian Hamiltonian with Lindblad operator but with $s = 0$}. In this case, the Hamiltonian $H$ and the Lindblad equation for the generalized density matrix $\rho^{}_{\rm G}$ are summarized as
\begin{equation}
H = \left( \begin{matrix} r e^{{\rm i} \varphi} & 0 \cr 0 & r e^{-{\rm i} \varphi} \end{matrix} \right) \quad \Leftrightarrow \quad \left( \begin{matrix} \dot{\rho}^{}_{{\rm G},0} \cr \dot{\rho}^{}_{{\rm G},1} \cr \dot{\rho}^{}_{{\rm G},2} \cr \dot{\rho}^{}_{{\rm G},3} \end{matrix} \right) = -2 \underbrace{\left( \begin{matrix} 0 & 0 & 0 & 0 \cr 0 & 0 & {\rm i} r \sin \varphi & 0 \cr 0 & -{\rm i} r \sin \varphi & \xi & 0 \cr 0 & 0 & 0 & \xi \end{matrix} \right)}_{R^{}_{\rm G}} \left( \begin{matrix} \rho^{}_{{\rm G},0} \cr \rho^{}_{{\rm G},1} \cr \rho^{}_{{\rm G},2} \cr \rho_{G,3} \end{matrix} \right) \,,
\end{equation}
where the non-Hermitian Hamiltonian $H$ is a complex diagonal matrix, and this case has a trivial metric. Since $R^{}_{\rm G}$ relates only the components $\rho^{}_{{\rm G},1}$, $\rho^{}_{{\rm G},2}$, and $\rho^{}_{{\rm G},3}$ of the generalized density matrix, the trace of $\rho^{}_{\rm G}(t)$ is equal to $2 \rho^{}_{{\rm G},0}(0) = 1$, which means that $\rho^{}_{{\rm G},0}(0) = 1/2$. In addition, the mixing matrix $A^{}_{\rm inv}$ is trivial, since the Hamiltonian is diagonal (although complex). Hence, the $+$ and $-$ states are identical to the $a$ and $b$ states. Notice that the eigenvectors are $(1,0)$ and $(0,1)$ and that the general formulas for the eigenvectors, i.e.~Eq.~(\ref{eq:neigenv}), cannot be used, since $s = 0$. The transition probabilities for both the $+$ and $-$ states and the $a$ and $b$ states are given by
\begin{align}
P_{++} = P_{--} = P_{aa} = P_{bb} &= \frac{1}{2} \left( 1 + e^{-2 \xi t} \right) \,, \label{eq:P++--} \\
P_{+-} = P_{-+} = P_{ab} = P_{ba} &= \frac{1}{2} \left( 1 - e^{-2 \xi t} \right) \,, \label{eq:P+--+}
\end{align}
where the Lindblad parameter $\xi > 0$ brings in dissipative effects. In the large time limit $t \to +\infty$, all transition probabilities approach $1/2$, indicating the complete loss of quantum coherence. In Fig.~\ref{fig:PTs=0}, choosing the value of the Lindblad parameter $\xi = 0.1$, we display the transition probabilities, where one can clearly observe the effect of damping.
\begin{figure}[t!]
\begin{center}
\includegraphics[width=0.5\textwidth]{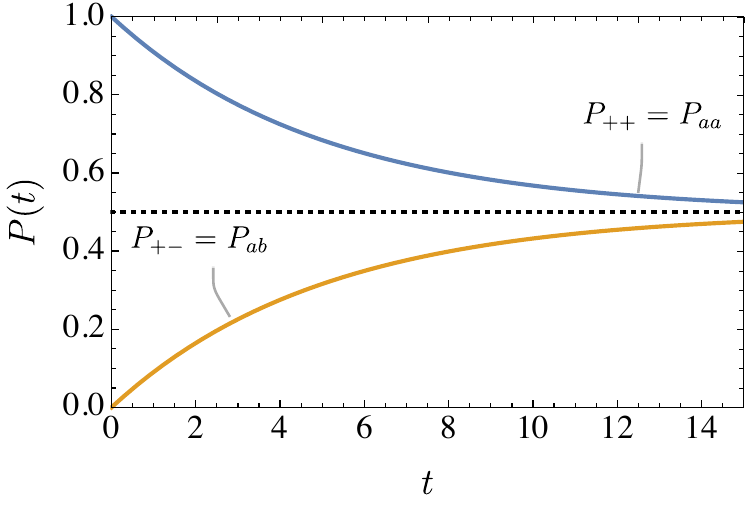}
\end{center}
\vspace{-5mm}
\caption{Transition probabilities for $+$ and $-$ states and $a$ and $b$ states in the case of a PT-symmetric non-Hermitian Hamiltonian with Lindblad term but with $s = 0$. Parameter value: $\xi = 0.1$.}
\label{fig:PTs=0}
\end{figure}

In order to have mixing, one can introduce an arbitrary rotation (see e.g.~Ref.~\cite{Ohlsson:2000mj}), but it is not necessary. In the case of mixing, we find the following formulas for the transition probabilities
\begin{align}
P_{\alpha\alpha}(\theta) = P_{\beta\beta}(\theta) &= \frac{1}{2} \left[ 1 + e^{-2 \xi t} \cos^22\theta + e^{-\xi t} \sin^22\theta \left( \cosh \widehat{\Omega} t + \frac{\xi}{\widehat{\Omega}} \sinh \widehat{\Omega} t \right) \right] \,, \label{eq:Paabb} \\
P_{\alpha\beta}(\theta) = P_{\beta\alpha}(\theta) &= \frac{1}{2} \left[ 1 - e^{-2 \xi t} \cos^22\theta - e^{-\xi t} \sin^22\theta \left( \cosh \widehat{\Omega} t + \frac{\xi}{\widehat{\Omega}} \sinh \widehat{\Omega} t \right) \right] \,, \label{eq:Pabba}
\end{align}
where $\widehat{\Omega} \equiv \sqrt{4 r^2 \sin^2 \varphi + \xi^2}$. It turns out that all transition probabilities in Eqs.~(\ref{eq:P++--})--(\ref{eq:Pabba}) are independent of both $r$ and $\varphi$, except from the transition probabilities in Eqs.~(\ref{eq:Paabb}) and (\ref{eq:Pabba}) with an arbitrary rotation by the mixing angle $\theta$, i.e.~$P_{\alpha\alpha}(\theta) = P_{\beta\beta}(\theta)$ and $P_{\alpha\beta}(\theta) = P_{\beta\alpha}(\theta)$. Obviously, the transition probabilities in Eqs.~(\ref{eq:Paabb}) and (\ref{eq:Pabba}) will reduce to those without mixing in Eqs.~(\ref{eq:P++--}) and (\ref{eq:P+--+}) by setting $\theta = 0$.

\item \label{case:PTr=0} {\it PT-symmetric (non-)Hermitian Hamiltonian with Lindblad operator but with $r = 0$}.  In this case, the Hamiltonian $H$ is a real off-diagonal matrix and actually Hermitian, i.e.~$H^\dagger = H$, so the Lindblad equation for the ordinary density matrix $\rho(t)$ and that for the generalized density matrix $\rho^{}_{\rm G}(t)$ are the same. Namely, the metric is trivial $\eta = \mathbb{1}^{}_2$. More explicitly, we have
\begin{equation}
H = \left( \begin{matrix} 0 & s \cr s & 0 \end{matrix} \right) \quad \Leftrightarrow \quad \left( \begin{matrix} \dot{\rho}^{}_{0} \cr \dot{\rho}^{}_{1} \cr \dot{\rho}^{}_{2} \cr \dot{\rho}^{}_{3} \end{matrix} \right) = -2 \underbrace{\left( \begin{matrix} 0 & 0 & 0 & 0 \cr 0 & 0 & 0 & 0 \cr 0 & 0 & \xi & s \cr 0 & 0 & -s & \xi \end{matrix} \right)}_{R} \left( \begin{matrix} \rho^{}_{0} \cr \rho^{}_{1} \cr \rho^{}_{2} \cr \rho^{}_{3} \end{matrix} \right) \,,
\end{equation}
where the ordinary density matrix $\rho(t)$ is used. It is worthwhile to point out that even the normalized density matrix $\rho^{}_{\rm N}(t)$ is applicable, since it is reduced to the ordinary density matrix $\rho(t)$ when the Hamiltonian is Hermitian, implying that $H^{}_- = 0$ and thus ${\cal N}[\rho^{}_{\rm N}(t)] = 0$ in Eq.~(\ref{eq:N}). The structure of $R$ is effectively a $2 \times 2$ matrix and relates only components $\rho^{}_2$ and $\rho^{}_3$ of the density matrix. This means that $\rho_0(0) = 1/2$, since the trace of the density matrix should be normalized to 1. In this case, we have the transition probabilities
\begin{align}
P_{++} = P_{--} &= 1 \,, \\
P_{+-} = P_{-+} &= 0 \,, \\
P_{aa} = P_{bb} &= \frac{1}{2} \left[ 1 + e^{-2 \xi t} \cos(2st) \right] \,, \\
P_{ab} = P_{ba} &= \frac{1}{2} \left[ 1 - e^{-2 \xi t} \cos(2st) \right] \,.
\end{align}
Note that there is no damping in the formulas for $P_{++} = P_{--}$ and $P_{+-} = P_{-+}$, i.e.~there is no effect of the Lindblad operator for these transition probabilities. Therefore, in Fig.~\ref{fig:PTr=0}, we plot the transition probabilities only for the $a$ and $b$ states, where the parameter values $s = 0.2$ and $\xi = 0.1$ have been chosen.
\begin{figure}[t!]
\begin{center}
\includegraphics[width=0.5\textwidth]{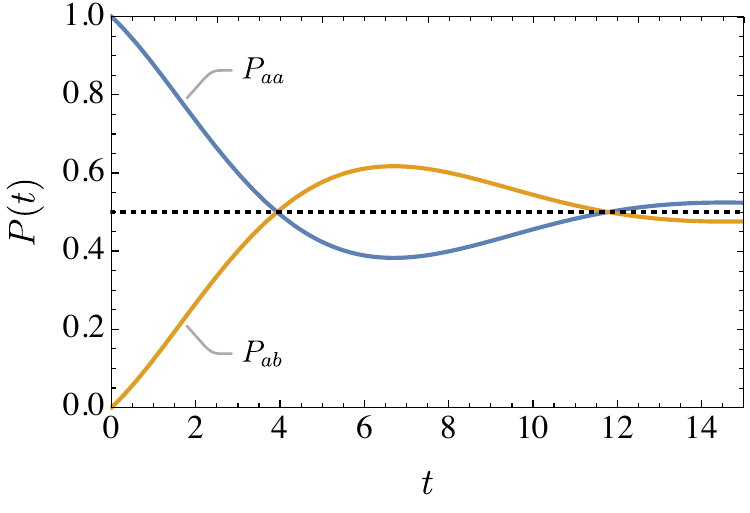}
\end{center}
\vspace{-5mm}
\caption{Transition probabilities for $a$ and $b$ states in the case of a PT-symmetric (non-)Hermitian Hamiltonian with Lindblad term but with $r = 0$. Parameter values: $s = 0.2$, $\xi = 0.1$.}
\label{fig:PTr=0}
\end{figure}

\item {\it Numerical example of a PT-symmetric non-Hermitian Hamiltonian with Lindblad operator}. This is the most non-trivial example with all relevant parameters being non-zero. The Hamiltonian $H$ and the Lindblad equation for the generalized density matrix $\rho^{}_{\rm G}(t)$ are
\begin{equation}
H = \left( \begin{matrix} r e^{{\rm i} \varphi} & s \cr s & r e^{-{\rm i} \varphi} \end{matrix} \right) \quad \Leftrightarrow \quad \left( \begin{matrix} \dot{\rho}^{}_{{\rm G},0} \cr \dot{\rho}^{}_{{\rm G},1} \cr \dot{\rho}^{}_{{\rm G},2} \cr \dot{\rho}^{}_{{\rm G},3} \end{matrix} \right) = -2 \underbrace{\left( \begin{matrix} 0 & 0 & 0 & 0 \cr 0 & 0 & {\rm i} r \sin \varphi & 0 \cr 0 & -{\rm i} r \sin \varphi & \xi & s \cr 0 & 0 & -s & \xi \end{matrix} \right)}_{R^{}_{\rm G}} \left( \begin{matrix} \rho^{}_{{\rm G},0} \cr \rho^{}_{{\rm G},1} \cr \rho^{}_{{\rm G},2} \cr \rho^{}_{{\rm G},3} \end{matrix} \right) \,,
\end{equation}
where $r = 0.1$, $s = 0.2$, $\varphi = \pi/3$, and $\xi = 0.1$ are adopted for numerical computations. This case has a non-Hermitian Hamiltonian and a non-trivial metric. Therefore, the generalized density matrix should be used. The structure of $R^{}_{\rm G}$ relates components $\rho^{}_{{\rm G},1}$, $\rho^{}_{{\rm G},2}$, and $\rho^{}_{{\rm G},3}$ of the generalized density matrix, and thus, the trace of the generalized density matrix is equal to $2 \rho^{}_{{\rm G},0}(0) = 1$, which means that $\rho^{}_{{\rm G},0}(0) = 1/2$. However, in order for the transition probabilities to be restricted between 0 and 1, the normalized generalized density matrix must be used after mixing with $A_{\rm inv}$, since the completeness relation changes due to the metric being non-trivial. In Fig.~\ref{fig:numerical}, we present the exact numerical results of the transition probabilities for the $a$ and $b$ states. Note that these transition probabilities are, of course, the same as those presented with solid curves in the right panel in Fig.~\ref{fig:approx}, since we use the same parameter values.
\begin{figure}[t!]
\begin{center}
\begin{tabular}{cc}
\includegraphics[width=0.48\textwidth]{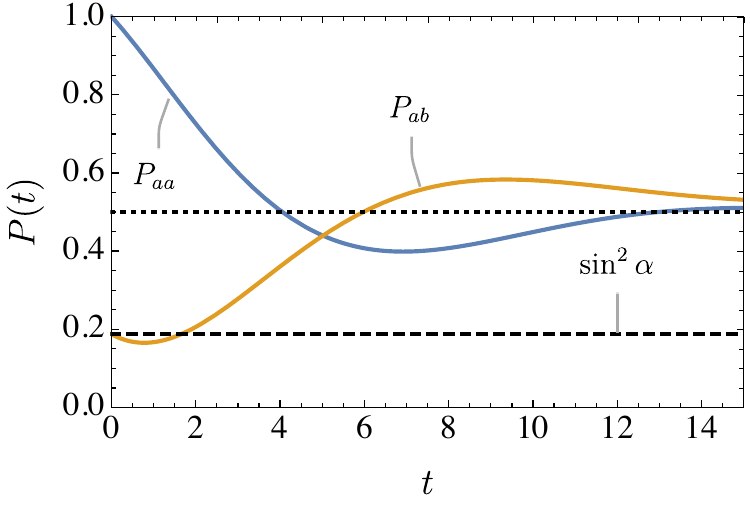} & \includegraphics[width=0.48\textwidth]{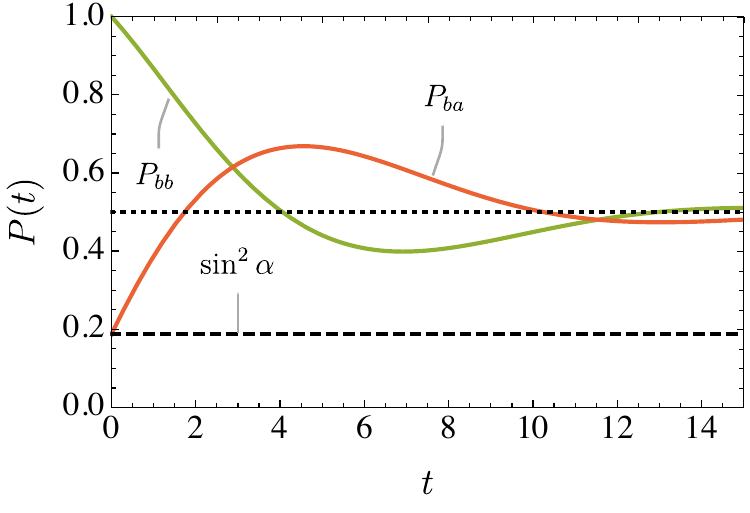}\\
\end{tabular}
\end{center}
\vspace{-5mm}
\caption{Numerical example of a PT-symmetric non-Hermitian Hamiltonian with Lindblad term. Transition probabilities for $a$ and $b$ states. {\it Left:} $P_{aa}$ and $P_{ab}$. {\it Right:} $P_{bb}$ and $P_{ba}$. Parameter values: $r = 0.1$, $s = 0.2$, $\varphi = \pi/3$, $\xi = 0.1$.}
\label{fig:numerical}
\end{figure}
\end{enumerate}

\subsection{The Lindblad Term with $\mbox{\boldmath$B = 0$}$}

\subsubsection{General Discussion}

Let us investigate the case for which the structure of the Lindblad operator is changed from being described by the assumption that the Lindblad parameter $A = 0$ to a different assumption that $B = 0$, which means that the other Lindblad parameters are $A = C$ and $D = E = F = 0$. Denoting the only non-zero Lindblad parameter by $\zeta$, we have $A = C = \zeta$. The explicit expression for $R^{}_{\rm G}(\phi = 0, B = 0)$ has been given in Eq.~(\ref{eq:RGB}).

\begin{figure}[t!]
\begin{center}
\begin{tabular}{cc}
\includegraphics[width=0.48\textwidth]{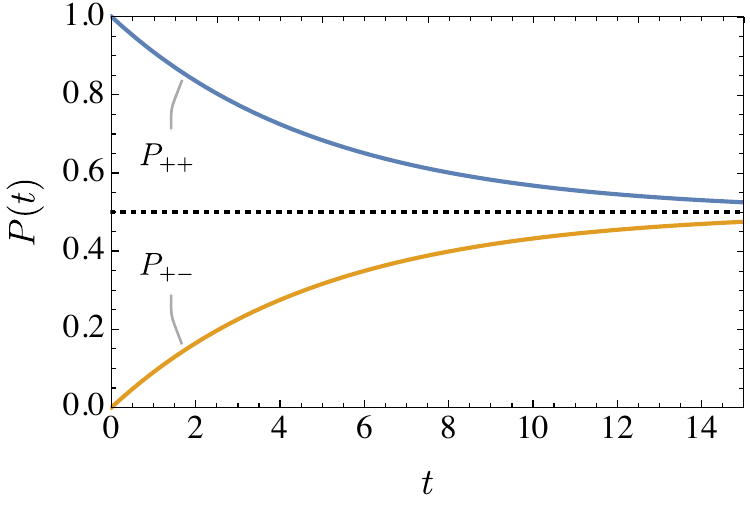} & \includegraphics[width=0.48\textwidth]{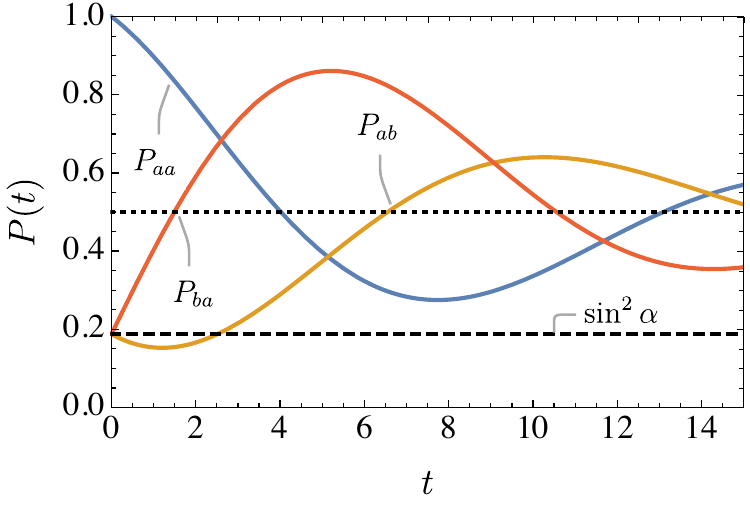}\\
\end{tabular}
\end{center}
\vspace{-5mm}
\caption{Exact transition probabilities as functions of time $t$ for the general PT-symmetric non-Hermitian Hamiltonian including the Lindblad term with $B = 0$. {\it Left:} Transition probabilities for $+$ and $-$ states. {\it Right:} Transition probabilities for $a$ and $b$ states. Parameter values: $r = 0.1$, $s = 0.2$, $\varphi = \pi/3$, $\zeta = 0.1$.}
\label{fig:exact}
\end{figure}

In comparison to the case $A = 0$, the case $B = 0$ turns out to be much easier to handle computationally with the method described in Subsec.~\ref{sec:LindbladA}. In fact, in the case $B = 0$, all transition probabilities can be computed as simple closed-form expressions. Using the described method, we first derive exact formulas for the transition probabilities between the $+$ and $-$ states and obtain
\begin{align}
P_{++}(t) = P_{--}(t) &= \frac{1}{2} \left( 1 + e^{-2 \zeta t} \right) \,, \label{eq:P++exact} \\
P_{+-}(t) = P_{-+}(t) &= \frac{1}{2} \left( 1 - e^{-2 \zeta t} \right) \,, \label{eq:P+-exact}
\end{align}
where the Lindblad parameter $\zeta > 0$, causing the damping of the transition probabilities as shown in the left panel of Fig.~\ref{fig:exact}. For our numerical computations, we choose $r = 0.1$, $s = 0.2$, $\varphi = \pi/3$, and $\zeta = 0.1$. In the large time limit $t \to +\infty$, all transition probabilities for the $+$ and $-$ states approach $1/2$.

Next, we proceed to the transition probabilities between the $a$ and $b$ states. The survival probabilities are found to be
\begin{equation}
P^{}_{aa}(t) = P^{}_{bb}(t) = \frac{1}{2} \left\{ 1 + e^{-\zeta t} \left[ \cos \Xi t - \frac{\zeta}{\Xi } \left( 1 - 2 \frac{r^2 \sin^2 \varphi}{s^2} \right) \sin \Xi t \right] \right\} \,, \label{eq:Paaexact}
\end{equation}
whereas the transition probabilities are given by
\begin{align}
P^{}_{ab}(t) &= \frac{1}{2} \left\{ 1 - e^{-\zeta t} \left[ \left( 1 - 2 \frac{r^2 \sin^2 \varphi}{s^2} \right) \cos \Xi t - \frac{\zeta - 4 r \sin \varphi (1 - r^2 \sin^2 \varphi/s^2)}{\Xi} \sin \Xi t \right] \right\} \,, \\
P^{}_{ba}(t) &= \frac{1}{2} \left\{ 1 - e^{-\zeta t} \left[ \left( 1 - 2 \frac{r^2 \sin^2 \varphi}{s^2} \right) \cos \Xi t - \frac{\zeta + 4 r \sin \varphi (1 - r^2 \sin^2 \varphi/s^2)}{\Xi} \sin \Xi t \right] \right\} \,, \label{eq:Pbaexact}
\end{align}
where $\Xi \equiv 2 \sqrt{s^2 - r^2 \sin^2 \varphi - \zeta^2/4}$. It should be noted that $P_{aa} (t = 0) = P_{bb}(t = 0) = 1$, whereas $P_{ab}(t=0) = P_{ba}(t=0) = r^2 \sin^2 \varphi/s^2 = \sin^2 \alpha \neq 0$ (in general), which is an effect of the non-trivial metric. Only if $r \sin \varphi = 0$ or $r \sin \varphi = \pm s$, then $P^{}_{ab}(t) = P^{}_{ba}(t)$, which means that $\sin \alpha = 0$ or $\sin \alpha = \pm 1$. However, in the limit $t \to +\infty$, it holds that $\lim_{t\to\infty} P^{}_{aa}(t) = \lim_{t\to\infty} P^{}_{bb}(t) = \lim_{t\to\infty} P^{}_{ab}(t) = \lim_{t\to\infty} P^{}_{ba}(t) = 1/2$. Thus, for large $t$, probability is conserved, i.e.~$P_{aa}(t) + P_{ab}(t) = 1$ and $P_{bb}(t) + P_{ba}(t) = 1$. In the right panel of Fig.~\ref{fig:exact}, we present the transition probabilities for the $a$ and $b$ states given by Eqs.~(\ref{eq:Paaexact})--(\ref{eq:Pbaexact}), using the parameter values $r = 0.1$, $s = 0.2$, $\varphi = \pi/3$, and $\zeta = 0.1$. The difference between $P^{}_{ab}(t)$ and $P^{}_{ba}(t)$ can be easily calculated as
\begin{equation}
P^{}_{ab}(t) - P^{}_{ba}(t) = - \frac{4 r \sin \varphi (1 - r^2 \sin^2 \varphi/s^2)}{\Xi}e^{-\zeta t}\sin \Xi t \,,
\end{equation}
which turns out to be an oscillatory sine function with a damping factor and vanishes in the limits of $t = 0$ and $t \to +\infty$. Compare the discussion on Eqs.~(\ref{eq:PaaPab}) and (\ref{eq:PbbPba}) for the case $A = 0$.

\subsubsection{Specific Examples with $\mbox{\boldmath$B = 0$}$}

Although it is possible to obtain exact analytical results for the Lindblad operator with $B = 0$, we briefly discuss a few specific examples with the same values of the relevant parameters as in the case with $A = 0$. Such comparative investigation should be helpful for us to understand the similarity and difference between these two cases.
\begin{enumerate}
\item {\it Two-flavor neutrino oscillations}. In this case, we obtain the transition probabilities
\begin{align}
P^{}_{++} = P^{}_{--} &= \frac{1}{2} \left( 1 + e^{-2 \zeta t} \right) \,, \\
P^{}_{+-} = P^{}_{-+} &= \frac{1}{2} \left( 1 - e^{-2 \zeta t} \right) \,, \\
P^{}_{aa} = P^{}_{bb} &= \frac{1}{2} \left[ 1 + e^{-2 \zeta t} \cos^22\theta + e^{-\zeta t} \sin^22\theta \left( \cos \widetilde\Omega t - \frac{\zeta}{\widetilde\Omega} \sin \widetilde\Omega t \right) \right] \,, \\
P^{}_{ab} = P^{}_{ba} &= \frac{1}{2} \left[ 1 - e^{-2 \zeta t} \cos^22\theta - e^{-\zeta t} \sin^22\theta \left( \cos \widetilde\Omega t - \frac{\zeta}{\widetilde\Omega} \sin \widetilde\Omega t \right) \right] \,,
\end{align}
where $\widetilde\Omega \equiv \sqrt{4 \omega^2 - \zeta^2}$. In comparison to special case~\ref{case:2fneuosc} for $A = 0$, the signs of the Lindblad parameters in front of the $\sin$ terms are opposite due to different positions of the Lindblad parameters in $R^{}_{\rm G}(\phi = 0, A = 0)$ and $R^{}_{\rm G}(\phi = 0, B = 0)$, respectively. At all times $t$, conservation of probability holds for both sets of transition probabilities, i.e.~$P^{}_{++} + P^{}_{+-} = 1$, $P^{}_{--} + P^{}_{-+} = 1$ and $P^{}_{aa} + P^{}_{ab} = 1$, $P^{}_{bb} + P^{}_{ba} = 1$. The formulas for $P^{}_{aa} = P^{}_{bb}$ and $P^{}_{ab} = P^{}_{ba}$ should be compared with the ones in Ref.~\cite{Benatti:2000ph}. In Fig.~\ref{fig:2fneuoscB}, we plot the formulas for the transition probabilities in the two-flavor neutrino oscillation case with a specific choice for the parameter values $\omega = 0.2$, $\zeta = 0.1$, and $\theta = \pi/3$, and for which one can find $\widetilde{\Omega} \approx 0.39$.
\begin{figure}[t!]
\begin{center}
\begin{tabular}{cc}
\includegraphics[width=0.48\textwidth]{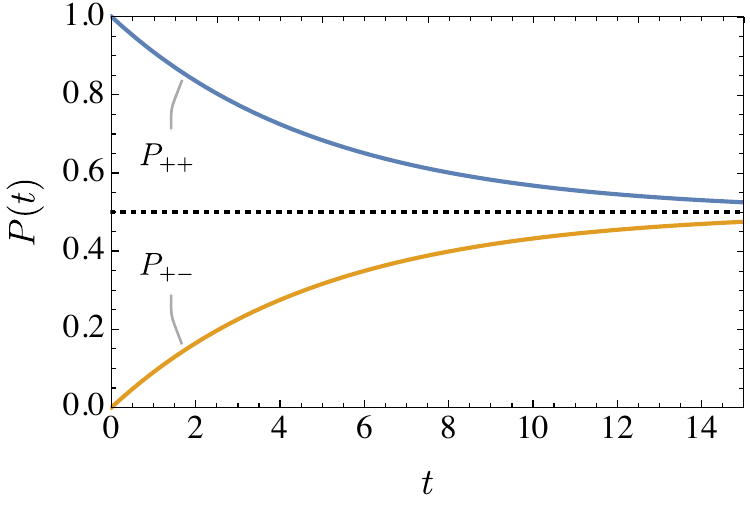} & \includegraphics[width=0.48\textwidth]{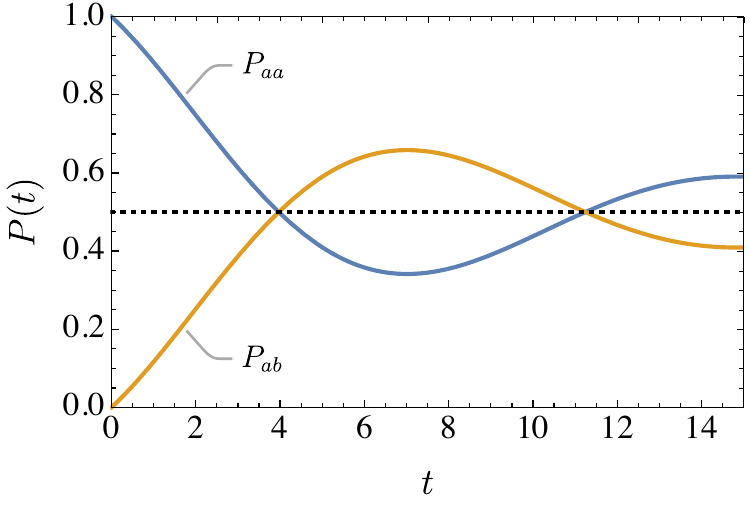}\\
\end{tabular}
\end{center}
\vspace{-5mm}
\caption{Two-flavor neutrino oscillations. {\it Left:} Transition probabilities for ``mass'' states. {\it Right:} Transition probabilities for ``flavor'' states. Parameter values: $\omega = 0.2$, $\zeta = 0.1$, $\widetilde{\Omega} \approx 0.39$, $\theta = \pi/3$.}
\label{fig:2fneuoscB}
\end{figure}

\item {\it Special case $\zeta = 0$}. In this case, the Lindblad term is absent. Since $R$ is the same in Eqs.~(\ref{eq:RGA}) and (\ref{eq:RGB}) if $\xi = 0$ and $\zeta = 0$, this case reproduces all results in special case~\ref{case:PTwoLindblad} for $A = 0$. Thus, it is unnecessary to repeat the results here.

\item {\it Special case $s=0$}. This case produces similar results as in special case~\ref{case:PTs=0} for $A = 0$, namely
\begin{align}
P_{++} = P_{--} = P_{aa} = P_{bb} &= \frac{1}{2} \left( 1 + e^{-2 \zeta t} \right) \,, \\
P_{+-} = P_{-+} = P_{ab} = P_{ba} &= \frac{1}{2} \left( 1 - e^{-2 \zeta t} \right) \,,
\end{align}
for the transition probabilities. As mentioned in special case~\ref{case:PTs=0} for $A = 0$, the Hamiltonian is diagonal, and thus, the $+$ and $-$ states and the $a$ and $b$ states coincide with each other, implying no mixing at all. If one introduces an arbitrary rotation by the angle $\theta$, the transition probabilities for the new $a$ and $b$ states are
\begin{align}
P_{\alpha\alpha}(\theta) = P_{\beta\beta}(\theta) &= \frac{1}{2} \left[ 1 + e^{-2 \zeta t} \cos^22\theta + e^{-\zeta t} \sin^22\theta \left( \cosh \varpi t - \frac{\zeta}{\varpi} \sinh \varpi t \right) \right] \,, \\
P_{\alpha\beta}(\theta) = P_{\beta\alpha}(\theta) &= \frac{1}{2} \left[ 1 - e^{-2 \zeta t} \cos^22\theta - e^{-\zeta t} \sin^22\theta \left( \cosh \varpi t - \frac{\zeta}{\varpi} \sinh \varpi t \right) \right] \,,
\end{align}
where $\varpi \equiv \sqrt{4 r^2 \sin^2 \varphi + \zeta^2}$. However, compared to the counterparts in special case~\ref{case:PTs=0} for $A = 0$, the signs of the Lindblad parameters in front of the $\sinh$ terms are opposite due to different positions of the Lindblad parameters in $R^{}_{\rm G}(\phi = 0, A = 0)$ and $R^{}_{\rm G}(\phi = 0, B = 0)$, respectively. At all times $t$, conservation of probability holds for all three sets of transition probabilities, i.e.~$P_{++} + P_{+-} = 1$, $P_{--} + P_{-+} = 1$, $P_{aa} + P_{ab} = 1$, $P_{bb} + P_{ba} = 1$, and $P_{\alpha\alpha}(\theta) + P_{\alpha\beta}(\theta) = 1$, $P_{\beta\beta}(\theta) + P_{\beta\alpha}(\theta) = 1$. In Fig.~\ref{fig:case_r}, we display the transition probabilities for the $+$ and $-$ states and the $a$ and $b$ states, which are the same as those in Fig.~\ref{fig:PTs=0}.
\begin{figure}[t!]
\begin{center}
\includegraphics[width=0.5\textwidth]{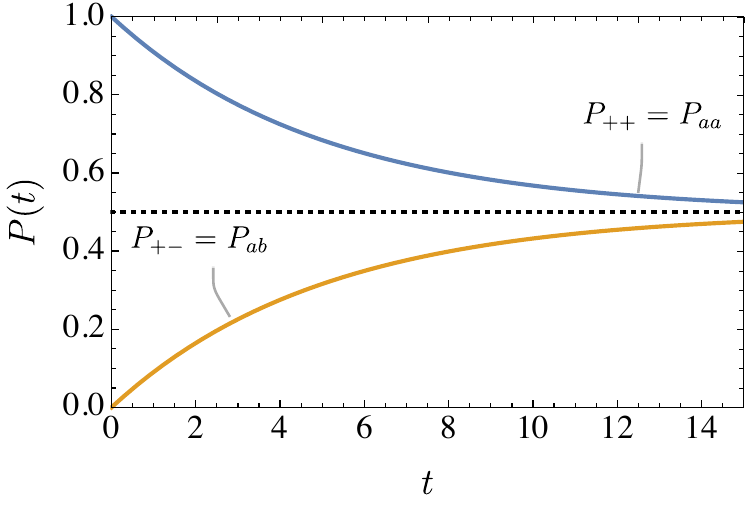}
\end{center}
\vspace{-5mm}
\caption{Special case $s=0$ for $B = 0$. Transition probabilities for $+$ and $-$ states and $a$ and $b$ states. Parameter value: $\zeta = 0.1$.}
\label{fig:case_r}
\end{figure}

\item {\it Special case $r=0$}. Inserting $r = 0$ into Eqs.~(\ref{eq:P++exact})--(\ref{eq:Pbaexact}), we find the transition probabilities
\begin{align}
P^{}_{++} = P^{}_{--} &= \frac{1}{2} \left( 1 + e^{-2 \zeta t} \right) \,, \\
P^{}_{+-} = P^{}_{-+} &= \frac{1}{2} \left( 1 - e^{-2 \zeta t} \right) \,, \\
P^{}_{aa} = P^{}_{bb} &= \frac{1}{2} \left[ 1 + e^{-\zeta t} \left( \cos \Sigma t - \frac{\zeta}{\Sigma} \sin \Sigma t \right) \right] \,, \\
P^{}_{ab} = P^{}_{ba} &= \frac{1}{2} \left[ 1 - e^{-\zeta t} \left( \cos \Sigma t - \frac{\zeta}{\Sigma} \sin \Sigma t \right) \right] \,,
\end{align}
where $\Sigma \equiv 2 \sqrt{s^2 - \zeta^2/4}$. Note that in the limit $r \to 0$, the non-trivial metric $\eta$ encoded in Eqs.~(\ref{eq:P++exact})--(\ref{eq:Pbaexact}) goes to the trivial metric $\eta \to \mathbb{1}^{}_2$. Furthermore, one can observe the differences compared to special case~\ref{case:PTr=0} for $A = 0$. At all times $t$, conservation of probability holds for both sets of transition probabilities, i.e.~$P_{++} + P_{+-} = 1$, $P_{--} + P_{-+} = 1$ and $P_{aa} + P_{ab} = 1$, $P_{bb} + P_{ba} = 1$. In Fig.~\ref{fig:case_s}, we show the transition probabilities for the $+$ and $-$ states in the left panel and those for the $a$ and $b$ states in the right panel.
\begin{figure}[t!]
\begin{center}
\begin{tabular}{cc}
\includegraphics[width=0.48\textwidth]{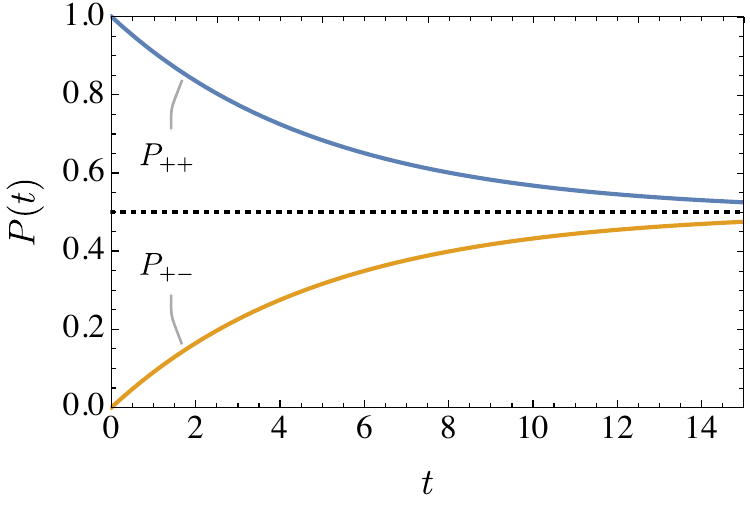} & \includegraphics[width=0.48\textwidth]{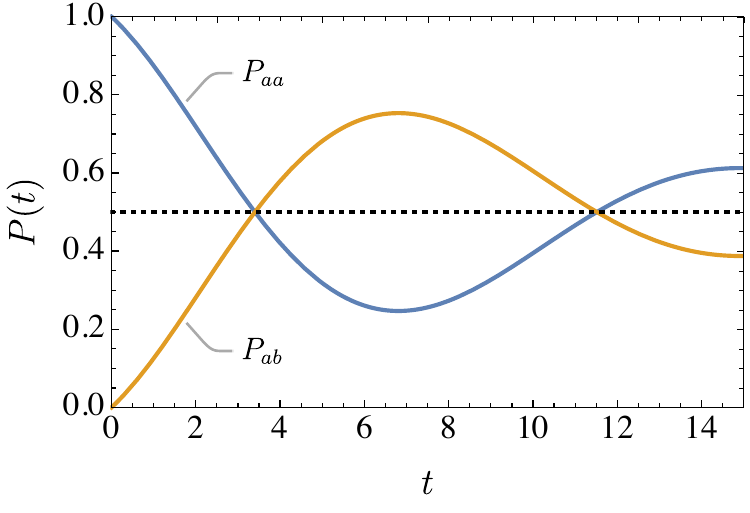}\\
\end{tabular}
\end{center}
\vspace{-5mm}
\caption{Special case $r=0$ for $B = 0$. {\it Left:} Transition probabilities for $+$ and $-$ states. {\it Right:} Transition probabilities for $a$ and $b$ states. Parameter values: $s = 0.2$, $\zeta = 0.1$, $\Sigma \approx 0.39$.}
\label{fig:case_s}
\end{figure}
\end{enumerate}

\section{Summary and Conclusions}
\label{sec:summary}

In this work, we have developed the density matrix formalism for PT-symmetric non-Hermitian Hamiltonians, and obtained the Lindblad equation for the generalized density matrix. Moreover, we have applied this formalism with the Lindblad equation to calculate the transition probabilities of $+$ and $-$ states and those of $a$ and $b$ states in the two-level quantum system. For a general two-level quantum system, the $+$ and $-$ states should be understood as the energy eigenstates of the system in question, whereas the $a$ and $b$ states are the initially prepared quantum states as coherent superpositions of the two energy eigenstates. Our main results are summarized as follows.

First, we have performed a comparative study of the generalized density matrix $\rho^{}_{\rm G}(t) \equiv \rho(t)\eta$ and the normalized density matrix $\rho^{}_{\rm N}(t) \equiv \rho(t)/{\rm tr}\left[\rho(t)\right]$ for non-Hermitian Hamiltonians $H$. In particular, we have generalized their evolution equations to the case where the Lindblad term is present. The Lindblad equation is very useful to account for the dissipative effects due to the interaction of the quantum system in question with the external environment. In the case of generalized density matrix $\rho^{}_{\rm G}\equiv \rho(t)\eta$, where $\eta$ is the metric such that the pseudo-Hermiticity $H^\dagger = \eta H \eta^{-1}$ is fulfilled, the Lindblad equation has been first derived for the pseudo-Hermitian Hamiltonian in Refs.~\cite{Scolarici:2006, Scolarici:2007}. However, we have pointed out that the condition of pseudo-Hermiticity should be imposed on the Lindblad operators $L^{}_j$, namely, $L^\dagger_j = \eta L^{}_j \eta^{-1}$. The implications of the pseudo-Hermiticity for the parameters involved in the Lindblad operators have been examined. In the case of normalized density matrix $\rho^{}_{\rm N}$, the Liouville--von Neumann equation contains an extra term, which turns out to be quadratic in the elements of the normalized density matrix $\rho^{}_{\rm N}(t)$~\cite{Sergi:2013}. Such a non-linearity hinders the generalization to the case with the Lindblad term, which is usually derived under the assumption of linearity.

Second, based on the Lindblad equation for the generalized density matrix, we have then calculated the transition probabilities of both $+$ and $-$ states and $a$ and $b$ states in the two-level system with PT-symmetric non-Hermitian Hamiltonians. We have first proposed a new parametrization of the Lindblad operators, fulfilling pseudo-Hermiticity, and arrived at two interesting cases of either $A = 0$ or $B = 0$, where $A$ and $B$ are two parameters in the Lindblad term. It is worth emphasizing that exact analytical formulas for the transition probabilities have been found in Eqs.~(\ref{eq:Paaexact})--(\ref{eq:Pbaexact}) in the most general case of $B = 0$, where only approximate results in the case of $A = 0$ can be obtained. Several concrete examples have then been presented, for which both analytical and numerical results of the transition probabilities are computed and compared with the existing results in the literature whenever a comparison is possible. These typical examples are instructive to clarify the main features of the Lindblad equation for PT-symmetric non-Hermitian Hamiltonians.

Finally, in the near future on the experimental side, it would be interesting to see whether the Lindblad equation derived for the generalized density matrix for PT-symmetric non-Hermitian Hamiltonians can be applied to realistic systems, such as those in optics and electronics, where the PT-symmetry has been experimentally observed. On the theoretical side, we hope to explore the general derivation of the Lindblad equation from the requirements for Markov limit, pseudo-Hermiticity, linearity, and complete positivity as in ordinary quantum mechanics.

\section*{Acknowledgments}

T.O.~acknowledges support by the Swedish Research Council (Vetenskapsr{\aa}det) through contract No.~2017-03934 and the KTH Royal Institute of Technology for a sabbatical period at the University of Iceland. The work of S.Z.~was supported in part by the National Natural Science Foundation of China under grant No.~11775232 and No.~11835013, and by the CAS Center for Excellence in Particle Physics.

\section*{Data Availability Statement}

The data that support the findings of this study are available from the corresponding author upon reasonable request.

\appendix

\section{The Lindblad Term}
\label{app:lindbladterm}

In the main text, we encounter the expansions of the density matrix $\rho(t)$ and the Lindblad operators $L_j$ in terms of Pauli matrices ${\bm \sigma}$ and the identity matrix $\sigma^{}_0 = \mathbb{1}^{}_2$. For reference, we collect the details of the expansions in this appendix. A general density matrix and the Lindblad operators can be expressed as follows
\begin{equation}
\rho = \sum^3_{\mu = 0} \rho^{}_\mu \sigma^{}_\mu = \rho^{}_0 \sigma^{}_0 + \bm{\rho} \cdot {\bm \sigma} \,, \quad L^{}_j = \sum^3_{\mu = 0} L^\mu_j \sigma^{}_\mu = L^0_j \sigma^{}_0 + \bm{L}^{}_j \cdot {\bm \sigma} \,.
\end{equation}
The Lindblad term ${\cal L}[\rho(t)]$, appearing in the evolution equations of the ordinary density matrix $\rho(t)$, the generalized one $\rho^{}_{\rm G}(t)$, and the normalized one $\rho^{}_{\rm N}(t)$, is given by
\begin{equation}
{\cal L}[\rho(t)] = -\frac{1}{2} \sum^3_{j=1} \left[L^2_j \rho(t) + \rho(t) L^2_j \right] + \sum^3_{j=1} L^{}_j \rho(t) L^{}_j \,, \label{eq:Lrt}
\end{equation}
where $\rho(t)$ can also be replaced by $\rho^{}_{\rm G}(t)$ or $\rho^{}_{\rm N}(t)$. Nevertheless, instead of $L^{}_j = L^\dagger_j$, the pseudo-Hermiticity condition $L^{}_j = \eta L^{}_j \eta^{-1} = L^\ddag_j$ holds in the latter two cases. Noticing the identity
\begin{equation}
(\bm{L}^{}_j \cdot \bm{\sigma})^2 = \sum_{i, k = 1}^3 L^i_j L^k_j \sigma^{}_i \sigma^{}_k = \frac{1}{2} \sum_{i, k = 1}^3 L^i_j L^k_j (\sigma^{}_i \sigma^{}_k + \sigma^{}_k \sigma^{}_i) = \sum_{i, k = 1}^3 L^i_j L^k_j \delta^{}_{ik} \sigma^{}_0 = \bm{L}^2_j \sigma^{}_0 \,,
\end{equation}
one can write
\begin{align}
L^2_j \rho &= \left[ (L^0_j)^2 \sigma^{}_0 + 2L^0_j \bm{L}^{}_j \cdot \bm{\sigma} + \bm{L}^2_j \sigma^{}_0 \right] \left(\rho^{}_0 \sigma^{}_0 + \bm{\rho} \cdot \bm{\sigma} \right) \nonumber \\
&= \rho^{}_0 \left[(L^0_j)^2 + \bm{L}^2_j\right] \sigma^{}_0 + 2\rho^{}_0 L^0_j \bm{L}^{}_j \cdot \bm{\sigma} + \left[(L^0_j)^2 + \bm{L}^2_j\right] \bm{\rho}\cdot \bm{\sigma} + 2L^0_j (\bm{L}^{}_j \cdot \bm{\sigma}) (\bm{\rho}\cdot \bm{\sigma}) \,, \\
\rho L^2_j &= \left(\rho^{}_0 \sigma^{}_0 + \bm{\rho} \cdot \bm{\sigma} \right) \left[ (L^0_j)^2 \sigma^{}_0 + 2L^0_j \bm{L}^{}_j \cdot \bm{\sigma} + \bm{L}^2_j \sigma^{}_0 \right] \nonumber \\
&= \rho^{}_0 \left[(L^0_j)^2 + \bm{L}^2_j\right] \sigma^{}_0 + 2\rho^{}_0 L^0_j \bm{L}^{}_j \cdot \bm{\sigma} + \left[(L^0_j)^2 + \bm{L}^2_j\right] \bm{\rho}\cdot \bm{\sigma} + 2L^0_j (\bm{\rho}\cdot \bm{\sigma}) (\bm{L}^{}_j \cdot \bm{\sigma}) \,,
\end{align}
and thus, we obtain
\begin{align}
-\frac{1}{2} \left(L^2_j \rho + \rho L^2_j \right) = &- \rho^{}_0 \left[(L^0_j)^2 + \bm{L}^2_j\right] \sigma^{}_0 - 2\rho^{}_0 L^0_j \bm{L}^{}_j \cdot \bm{\sigma} - \left[(L^0_j)^2 + \bm{L}^2_j\right] \bm{\rho}\cdot \bm{\sigma} \nonumber \\
&- L^0_j (\bm{\rho}\cdot \bm{\sigma}) (\bm{L}^{}_j \cdot \bm{\sigma}) - L^0_j (\bm{L}^{}_j \cdot \bm{\sigma}) (\bm{\rho}\cdot \bm{\sigma}) \,. \label{eq:L2r+rL2}
\end{align}
On the other hand, we have
\begin{align}
L^{}_j \rho L^{}_j &= \left( L^0_j \sigma^{}_0 + \bm{L}^{}_j \cdot \bm{\sigma} \right) \left(\rho^{}_0 \sigma^{}_0 + \bm{\rho} \cdot \bm{\sigma} \right) \left( L^0_j \sigma^{}_0 + \bm{L}^{}_j \cdot \bm{\sigma} \right) \nonumber \\
&= \rho^{}_0 \left[(L^0_j)^2 + \bm{L}^2_j\right] \sigma^{}_0 +2\rho^{}_0 L^0_j \bm{L}^{}_j \cdot \bm{\sigma} + (L^0_j)^2 \bm{\rho}\cdot \bm{\sigma} + L^0_j (\bm{\rho}\cdot \bm{\sigma}) (\bm{L}^{}_j \cdot \bm{\sigma}) \nonumber \\
&+ L^0_j (\bm{L}^{}_j \cdot \bm{\sigma}) (\bm{\rho}\cdot \bm{\sigma}) + (\bm{L}^{}_j \cdot \bm{\sigma}) (\bm{\rho}\cdot \bm{\sigma}) (\bm{L}^{}_j \cdot \bm{\sigma}) \,. \label{eq:LrL}
\end{align}
Inserting Eqs.~(\ref{eq:L2r+rL2}) and (\ref{eq:LrL}) into Eq.~(\ref{eq:Lrt}), the Lindblad term can be rewritten as
\begin{equation}
{\cal L}[\rho] = -\sum_{j = 1}^3 \bm{L}^2_j (\bm{\rho}\cdot \bm{\sigma}) + \sum_{j = 1}^3 (\bm{L}^{}_j \cdot \bm{\sigma}) (\bm{\rho}\cdot \bm{\sigma}) (\bm{L}^{}_j \cdot \bm{\sigma}) \,,
\end{equation}
which can be further simplified to
\begin{equation}
{\cal L}[\rho] = -2\sum_{j = 1}^3 \bm{L}^2_j (\bm{\rho}\cdot \bm{\sigma}) + 2\sum_{j = 1}^3 (\bm{\rho}\cdot \bm{L}^{}_j) (\bm{L}^{}_j \cdot \bm{\sigma}) \,,
\label{eq:appLindblad}
\end{equation}
by using the identity
\begin{align}
(\bm{L}^{}_j \cdot \bm{\sigma}) (\bm{\rho}\cdot \bm{\sigma}) (\bm{L}^{}_j \cdot \bm{\sigma}) &= \sum_{\ell,m,n}^3 L^\ell_j \rho^{}_m L^n_j \sigma^{}_\ell \sigma^{}_m \sigma^{}_n
= \sum_{\ell,m,n = 1}^3 L^\ell_j \rho^{}_m L^n_j \sigma^{}_\ell \left(2\delta^{}_{m n} - \sigma^{}_n \sigma^{}_m\right) \nonumber \\
&= - \bm{L}^2_j (\bm{\rho} \cdot \bm{\sigma}) + 2 (\bm{\rho}\cdot \bm{L}^{}_j) (\bm{L}^{}_j \cdot \bm{\sigma}) \,.
\end{align}
The final form of the Lindblad term in Eq.~(\ref{eq:appLindblad}) has been used in our discussion in the main text.

\section{Four-Dimensional Representation}
\label{app:4Drep}

In the derivation of the evolution equations for the generalized density matrix $\rho^{}_{\rm G}(t)$ and the normalized density matrix $\rho^{}_{\rm N}(t)$ in the $4\times 4$ matrix formalism, we have to take the traces of the Lindblad equation and those together with the Pauli matrices ${\bm \sigma}$. In this appendix, we will give some key details of calculation. For the ordinary density matrix $\rho(t)$ in Hermitian quantum mechanics and $\rho^{}_{\rm G}(t)$ in the non-Hermitian quantum mechanics under consideration, the Liouville--von Neumann equations in both cases contain similar commutators, i.e.~$-{\rm i}[H, \rho(t)]$ and $-{\rm i}[H, \rho^{}_{\rm G}(t)]$. Using Einstein's summation convention, the trace of these commutators can be calculated as
\begin{equation}
-{\rm i} {\, {\rm tr}}\left\{[H, \rho]\right\} = - {\rm i} {\, {\rm tr}}\left\{[{\bm H}\cdot {\bm \sigma}, {\bm \rho}\cdot {\bm \sigma}]\right\} = 2 \epsilon^{}_{ijk} {\rm tr}\left(\sigma^{}_i\right) H^{}_j \rho^{}_k = 0 \,, \label{eq:tr1}
\end{equation}
whereas their trace with the Pauli matrix $\sigma^{}_i$ is
\begin{equation}
-{\rm i}{\, {\rm tr}}\left\{[H, \rho]\sigma^{}_i\right\} = - {\rm i}{\, {\rm tr}}\left\{[{\bm H}\cdot {\bm \sigma}, {\bm \rho}\cdot {\bm \sigma}]\sigma^{}_i\right\} = 4 \epsilon^{}_{ijk} B^{}_j \rho^{}_k \,, \label{eq:tr2}
\end{equation}
where $\epsilon^{}_{ijk}$ is the three-dimensional Levi-Civita tensor. Recall that the Hamiltonian $H$ and the density matrix $\rho$ are expanded as $H = H^{}_\mu \sigma^{}_\mu$ and $\rho = \rho^{}_\mu \sigma^{}_\mu$, where summation over $\mu = 0, 1, 2, 3$ is implied. By implementing Eqs.~(\ref{eq:tr1}) and (\ref{eq:tr2}), one can immediately recast the Lindblad equation in Eq.~(\ref{eq:Glindbladh}) into the four-dimensional representation as
\begin{equation}
\frac{\partial}{\partial t} \left( \begin{matrix} \rho^{}_0 \cr \rho^{}_1 \cr \rho^{}_2 \cr \rho^{}_3 \end{matrix} \right) = \left[ - 2 \left( \begin{matrix} 0 & 0 & 0 & 0 \cr 0 & 0 & H^{}_3 & -H^{}_2 \cr 0 & -H^{}_3 & 0 & H^{}_1 \cr 0 & H^{}_2 & -H^{}_1 & 0 \end{matrix} \right) - 2 \left( \begin{matrix} 0 & 0 & 0 & 0 \cr 0 & A & D & E \cr 0 & D & B & F \cr 0 & E & F & C \end{matrix} \right) \right] \left( \begin{matrix} \rho^{}_0 \cr \rho^{}_1 \cr \rho^{}_2 \cr \rho^{}_3 \end{matrix} \right) \,,
\label{eq:Glindblad4}
\end{equation}
where the Lindblad parameters $A$, $B$, $C$, $D$, $E$, and $F$ have been given in Eqs.~(\ref{eq:A})--(\ref{eq:F}). It is obvious that $\partial \rho^{}_0/{\partial t} = 0$ applies to the coefficient $\rho^{}_0$ of the ordinary density matrix $\rho(t)$ and also to that of the generalized density matrix $\rho^{}_{\rm G}(t)$.

However, for the normalized density matrix $\rho^{}_{\rm N}(t)$, we have to deal with $-{\rm i}\left(H \rho^{}_{\rm N} - \rho^{}_{\rm N} H^\dagger\right)$, where $H^\dagger$ is present as well. In the most general case, all coefficients $H^{}_\mu$ (for $\mu = 0, 1, 2, 3$) in the expansion of $H$ are complex. Therefore, we have
\begin{align}
-{\rm i}{\, {\rm tr}} \left(H \rho^{}_{\rm N} - \rho^{}_{\rm N} H^\dagger\right) &= -{\rm i}{\, {\rm tr}}\left[(H^{}_0 - H^*_0)\rho^{}_0\right] - {\rm i}{\, {\rm tr}}\left[({\bm H}\cdot {\bm \sigma} - {\bm H}^*\cdot {\bm \sigma})({\bm \rho}\cdot {\bm \sigma}) \right] \nonumber \\
&= 4 \Im(H^{}_0) \rho^{}_0 + 4 \Im(H^{}_i) \rho^{}_i \,.
\end{align}
Furthermore, if the trace is taken together with the Pauli matrix $\sigma^{}_i$, then we obtain
\begin{equation}
-{\rm i}{\, {\rm tr}} \left[(H \rho^{}_{\rm N} - \rho^{}_{\rm N} H^\dagger)\sigma^{}_i\right] = 4 \Im(H^{}_0) \rho^{}_i + 4 \Im(H^{}_i) \rho^{}_0 + 4 \epsilon^{}_{ijk} \Re(H^{}_j) \rho^{}_k \,.
\end{equation}
Since there is an extra term in Eq.~(\ref{eq:N}) in the Lindblad equation for the normalized density matrix $\rho^{}_{\rm N}(t)$, we should investigate the traces involving this term. The trace of the first term on the right-hand side of Eq.~(\ref{eq:N}) can be found as
\begin{align}
{\rm tr}\left(\rho^{}_{\rm N} \right) \cdot 2{\rm i}\, {\rm tr}\left[H^{}_- \rho^{}_{\rm N}(t)\right] &= 2{\rm i}\rho^{}_0 {\, {\rm tr}} \left[(H^{}_0 - H^*_0)\rho^{}_0\right] + 2{\rm i}\rho^{}_0 {\, {\rm tr}}\left[({\bm H} \cdot {\bm \sigma} - {\bm H}^* \cdot {\bm \sigma}) ({\bm \rho}\cdot {\bm \sigma})\right] \nonumber \\
&= -4 \rho^{}_0 \left[\Im(H^{}_0) \rho^{}_0 + \Im(H^{}_i) \rho^{}_i \right] \,,
\end{align}
which turns out to be non-linear in $\rho^{}_\mu$ (for $\mu = 0, 1, 2, 3$). When the trace is taken together with the Pauli matrix $\sigma^{}_i$, we obtain
\begin{align}
{\rm tr}\left(\rho^{}_{\rm N} \sigma^{}_i \right) \cdot 2{\rm i}\, {\rm tr}\left[H^{}_- \rho^{}_{\rm N}(t)\right] &= {\rm i}{\, {\rm tr}}\left(\rho \sigma^{}_i\right){\rm tr} \left[(H^{}_0 - H^*_0)\rho^{}_0\right] + {\rm i}{\, {\rm tr}}\left(\rho \sigma^{}_i\right) {\, {\rm tr}} \left[({\bm H} \cdot {\bm \sigma} - {\bm H}^* \cdot {\bm \sigma}) ({\bm \rho}\cdot {\bm \sigma})\right] \nonumber \\
&= -4 \rho^{}_i \left[\Im(H^{}_0) \rho^{}_0 + \Im(H^{}_j) \rho^{}_j \right] \,.
\end{align}
Thus, taking into account the first term in the right-hand side of Eq.~~(\ref{eq:N}), the four-dimensional representation of the Lindblad equation in Eq.~(\ref{eq:Nlindbladh}) for the normalized density matrix $\rho^{}_{\rm N}(t)$ becomes
\begin{align}
\frac{\partial}{\partial t} \left( \begin{matrix} \rho^{}_0 \cr \rho^{}_1 \cr \rho^{}_2 \cr \rho^{}_3 \end{matrix} \right) &= \left[ - 2 \left(\begin{matrix} 0 & -\Im(H^{}_1) & -\Im(H^{}_2) & -\Im(H^{}_3) \cr -\Im(H^{}_1) & 0 & \Re(H^{}_3) & -\Re(H^{}_2) \cr -\Im(H^{}_2) & -\Re(H^{}_3) & 0 & \Re(H^{}_1) \cr -\Im(H^{}_3) & \Re(H^{}_2) & -\Re(H^{}_1) & 0 \end{matrix}\right) - 2 \left( \begin{matrix} 0 & 0 & 0 & 0 \cr 0 & A & D & E \cr 0 & D & B & F \cr 0 & E & F & C \end{matrix} \right) \right] \left( \begin{matrix} \rho^{}_0 \cr \rho^{}_1 \cr \rho^{}_2 \cr \rho^{}_3 \end{matrix} \right) \nonumber \\
&- 2 \left( \begin{matrix} \rho^{}_0 & 0 & 0 & 0 \cr 0 & \rho^{}_1 & 0 & 0 \cr 0 & 0 & \rho^{}_2 & 0 \cr 0 & 0 & 0 & \rho^{}_3\end{matrix} \right) \left(\begin{matrix} \Im(H^{}_0) & \Im(H^{}_1) & \Im(H^{}_2) & \Im(H^{}_3) \cr \Im(H^{}_0) & \Im(H^{}_1) & \Im(H^{}_2) & \Im(H^{}_3) \cr \Im(H^{}_0) & \Im(H^{}_1) & \Im(H^{}_2) & \Im(H^{}_3) \cr \Im(H^{}_0) & \Im(H^{}_1) & \Im(H^{}_2) & \Im(H^{}_3) \end{matrix}\right) \left( \begin{matrix} \rho^{}_0 \cr \rho^{}_1 \cr \rho^{}_2 \cr \rho^{}_3 \end{matrix} \right) \,,
\label{eq:Nlindblad4}
\end{align}
where the last term is actually quadratic in $\rho^{}_\mu$ (for $\mu = 0, 1, 2, 3$). Such a non-linearity seems to be in contradiction with the assumption of linearity in the derivation of the Lindblad term. The second and third terms on the right-hand side of Eq.~(\ref{eq:N}) lead to further non-linear contributions. Therefore, only the generalized density matrix $\rho^{}_{\rm G}(t)$ and its Lindblad equation will be used in our calculation of transition probabilities.

\section{Procedure}
\label{app:procedure}

In this work, the transition probabilities for a two-level quantum system described by a PT-symmetric non-Hermitian Hamiltonian are calculated according to the following procedure:
\begin{itemize}
\item Given a Hamiltonian for a two-level system, construct the corresponding $2 \times 2$ density matrix $\rho$ expressed and parametrized in terms of the identity matrix and the three Pauli matrices.
\item Compute the eigenvalues and construct the properly normalized eigenvectors (see Ref.~\cite{Ohlsson:2019noy}) and then the metric and the initial generalized density matrices.
\item Determine the time evolution of the density matrix for the system. The time evolution is given by the Lindblad equation for the generalized density matrix $\rho_{\rm G}$.
\item Transform the $2 \times 2$ density matrix to a corresponding $4 \times 4$ matrix with the time evolution described by a Schr{\"o}dinger-like equation (with a $4 \times 4$ ``Hamiltonian'') for a vector with the four components of the original density matrix. Add ${\cal L}[\rho]$ to the $4 \times 4$ ``Hamiltonian'' and the result is denoted $R$, which encodes both contributions from the original Hamiltonian and the Lindblad term.
\item Solve the Schr{\"o}dinger-like equation by matrix exponentiation of $R$, which is denoted ${\cal M}(t)$ and leads to the time evolution for the vector of the density matrix components with the initial density matrix components as the initial condition.
\item Transform the vector of the density matrix components back to a $2 \times 2$ density matrix.
\item Compute the transition probabilities between the $+$ and $-$ states by ${\rm tr}\left[\rho^{}_{\rm G}(t) \rho^{}_{\rm G}(0)\right]$ for generalized density matrices.
\item Construct the mixing matrix $A^{}_{\rm inv}$ (based on the $+$ and $-$ states, known as ``mass'' states, see Ref.~\cite{Ohlsson:2019noy}), the normalized ``flavor'' eigenvectors (using $A^{}_{\rm inv}$), i.e.~the $a$ and $b$ states, and the initial ``flavor'' density matrices (using the normalized ``flavor'' eigenvectors and the metric).
\item Change the initial condition of the density matrix from the initial ``mass'' density matrices to the initial ``flavor'' density matrices. Note that the time evolution for the density matrix is the same in any basis, since ${\cal M}(t)$ only depends on time $t$ and the eigenvalues of $R$, which are the same in all bases.
\item Compute the transition probabilities between the $a$ and $b$ states by ${\rm tr} \left[\rho^{}_{\rm GN}(t) \rho^{}_{\rm GN}(0)\right]$. Note that if the metric is non-trivial, then, after mixing with $A_{\rm inv}$ (since the completeness relation changes due to the non-trivial metric), the normalized (generalized) density matrices must be used, so that the transition probabilities are restricted between 0 and 1.
\end{itemize}

\end{document}